\documentclass[journal]{IEEEtran}
\usepackage{amsmath,amsfonts,amssymb}
\newtheorem{definition}{Definition}
\usepackage{algorithmic,algorithm}
\usepackage{array}
\usepackage[caption=false,font=normalsize,labelfont=sf,textfont=sf]{subfig}
\usepackage{textcomp}
\usepackage{stfloats}
\usepackage{url}
\usepackage{verbatim}
\usepackage{graphicx}
\usepackage{cuted}
\usepackage{times,amsfonts}
\graphicspath{{pictures/},{Authorpictures/}}
\def\BibTeX{{\rm B\kern-.05em{\sc i\kern-.025em b}\kern-.08em
		T\kern-.1667em\lower.7ex\hbox{E}\kern-.125emX}}
\usepackage{balance}
\usepackage{cite}
\usepackage{float}
\usepackage{color}
\usepackage[hidelinks]{hyperref}
\begin{document}
	\title{Resource Allocation for RIS-Assisted Device-to-Device Communications in Heterogeneous Cellular Networks}
	\author{Shaoyou~Ao,
		Yong~Niu,~\IEEEmembership{Senior Member,~IEEE},
		Zhu~Han,~\IEEEmembership{Fellow,~IEEE},
		Bo~Ai,~\IEEEmembership{Fellow,~IEEE},
		Zhangdui~Zhong,~\IEEEmembership{Fellow,~IEEE},
		Ning~Wang,~\IEEEmembership{Member,~IEEE},
		and Yuanyuan Qiao
		
		\thanks{Copyright (c) 2015 IEEE. Personal use of this material is permitted. However, permission to use this material for any other purposes must be obtained from the IEEE by sending a request to pubs-permissions@ieee.org. This study was supported by National Key R\&D Program of China (2020YFB1806903); in part by the National Natural Science Foundation of China Grants 62231009, 62221001, 61801016, 61725101, 61961130391, and U1834210; in part by the Project of China Shenhua under Grant (GJNY-20-01-1); and in part by NSF CNS-2107216, CNS-2128368, CMMI-2222810, US Department of Transportation, Toyota and Amazon. (\emph{Corresponding author: Y. Niu.})}
		
		\thanks{S. Ao, Y. Niu, B. Ai, Z. Zhong,Y. Qiao are with the State Key Laboratory of Rail Traffic Control and Safety, Beijing Jiaotong University, Beijing 100044,	China(e-mails: 22120025@bjtu.edu.cn, niuy11@163.com, boai@bjtu.edu.cn, zhdzhong@bjtu.edu.cn, 21111050@bjtu.edu.cn).}
		
		\thanks{Z. Han is with the Department of Electrical and Computer Engineering, University of Houston, Houston, TX 77004 USA, and also with the Department of Computer Science and Engineering, Kyung Hee University, Seoul 446-701, South Korea (e-mail: hanzhu22@gmail.com).}
		
		\thanks{N. Wang is with the School of Information Engineering, Zhengzhou University, Zhengzhou 450001, China (e-mail: ienwang@zzu.edu.cn).}
	}	
	\maketitle
	\begin{abstract}
		In recent years, with the explosive growth of data traffic, communication base stations (BSs) need to serve more and more users. Offloading traffic from BSs has become an efficient way to reduce the burden on BSs.
		Device-to-Device (D2D) communications have emerged to improve spectrum utilization by reusing the frequency spectrum of the cellular frequency band.
		In the general environment, Heterogeneous Cellular Networks (HCNs) including millimeter wave (mm-wave) have appeared.
		Since the D2D link allows to share of spectrum resources with the cellular user, it will bring potential  interference to the cellular user.
		Fortunately, an emerging technology called Reconfigurable Intelligent Surface (RIS) can mitigate the severe interference caused by D2D links by shaping the incident beam and improving the multipath phase shift.
		In this paper, we study the resource allocation scheme to maximize the system sum rate, in the RIS-assisted single-cell heterogeneous D2D communication scenario.
		To solve the Block Coordinate Descent (BCD) problem, the problem of maximizing the sum rate is decomposed into three sub-problems.
		The resource allocation sub-problem is solved by a coalitional game method based on the game theory.
		The power allocation problem of the coalition converts the concave function into a convex optimization by mathematical transformation.  The problem is solved by the gradient descent method.
		The local search method is adopted to find the optimum for the phase conversion problem.
		Then iterate until the difference of sum rate is less than the threshold.
		The simulation results show that the designed algorithm is superior to other benchmark schemes in the literature.
		
	\end{abstract}
	
		
	\section{Introduction}\label{Sec1}
	\IEEEPARstart{I}{n recent} years, with the massive increase of smart mobile devices and smart terminals, and the popularization of multimedia services, the global mobile communication traffic will increase sharply. Cisco's report predicts that global mobile data traffic will increase sevenfold from 2017 to 2022, while mobile traffic will account for 20$\%$ of total IP traffic by 2022, of which 79$\%$ will be video traffic\cite{forecast2019cisco}. At the same time, the shortage of spectrum resources is also a major challenge currently facing. The millimeter wave(mm-wave) band can provide abundant bandwidth resources and provide higher network capacity. Several indoor wireless personal area networks (WPANs) or wireless local area networks (WLANs) standards for mm-wave have been proposed, such as ECMA-387\cite{ecma60ghz}, IEEE 802.15.3c\cite{5284444}, and IEEE 802.11ad. In Device-to-Device (D2D) communications, physically close users can communicate directly without going through a base station (BS).
	Due to the short physical distance, energy consumption can be reduced while ensuring user quality of service (QoS) requirements. Moreover, D2D links are allowed to share uplink spectrum with cellular users, which can also alleviate spectrum shortages.
	
	Heterogeneous cellular networks (HCNs) are cellular networks that include both cellular and mm-wave frequency bands that combine the different advantages offered by the two kinds of frequency bands. For example, cellular networks provide higher link reliability, and mm-wave networks have greater advantages in transmission rates. However, the existence of co-channel interference in the same frequency band will lead to the degradation of system performance. Therefore, how to manage the interference through mode switch and other methods has become one of the key issues to reduce the interference of D2D communication in HCNs. Furthermore, in the heterogeneous networks of cellular and mm-wave bands, one common problem is that mm-wave communication has larger propagation loss than cellular communication due to high carrier frequency. For example, the propagation loss in free space at 60 GHz is 28 dB higher than in the 2.4 GHz band\cite{singh2011interference}. Therefore, in the mm-wave band, users need more transmit power to ensure the reliability of transmission, and in addition,  to combat fading, the mm-wave band usually requires Line-of-Sight (LoS) communication\cite{9779354}.
	
	Fortunately, Reconfigurable Intelligent Surface (RIS) was proposed as a new low-cost technology to achieve high spectral and energy efficiency in wireless communication through low-cost reflective elements\cite{huang2019reconfigurable}. RIS is a two-dimensional ultra-thin reflective surface integrated with electronic circuits. It consists of many controllable and programmable positive intrinsic negative (PIN) diodes. The electromagnetic response of each element of the RIS can be tuned in a software-defined manner. On the other hand, RIS can make the wireless environment controllable and programmable in principle, thus it brings unprecedented new opportunities for improving the performance of wireless communication systems. In addition, RIS is lightweight and easy to install, and can be installed on the walls or ceilings of buildings, thereby integrating them into existing cellular networks and Wi-Fi systems without changing any BS hardware, hardware of access point, and user terminal\cite{wu2019beamforming}. It generates favorable beams by adjusting the phase shift of the signals, so as to limit the multipath effect\cite{zhang2021reconfigurable}. The reflected signals are superimposed constructively or destructively at the receiver to increase the required signal power or suppress co-channel interference, thereby improving communication performance. However, due to its passive feature \cite{tang2020wireless}, the reflection coefficient can only be taken between $\left[ {0,1} \right]$. How to design the phase shift of the RIS to enhance the received signal and reduce interference is an important and difficult problem. Second, power control is another feasible solution to mitigate interference among co-channel users. Therefore, using RIS to assist D2D communication in HCNs, as a potential method, can improve frequency band utilization, offload BS traffic, provide higher transmission rates, and provide users with a better experience.
	
	In this paper, we study an uplink RIS-assisted HCN in which D2D links can share the same frequency band as cellular links and an mm-wave frequency band, and RIS is deployed to manage interference. The optimization goal is to maximize the sum rate of the system while satisfying QoS constraints and transmission power constraints. The problem is a mixed integer nonlinear programming problem, which is challenging to solve. Because the constraints are coupled, we decompose the optimization problem into three sub-problems for alternate iterative solutions.
	
	The contributions of this paper are summarized as follows.	
	\begin{itemize}
		\item We establish an uplink RIS-assisted HCN, where the D2D link can share the same frequency band with the cellular frequency band or the mm-wave frequency band, and deploy RIS to provide reflection paths to enhance received power and mitigate interference.
		
		\item We formulate the system sum rate maximization problem by switching the working mode, optimizing the transmitted power and discrete phase shifts of RIS elements, subjected to the minimum signal-to-interference noise ratio (SINR) constraint, maximum transmitted power constraint, and phase shift constraints.
		
		\item Resource allocation adopts the algorithm of coalition game to solve the sub-problem of working mode switching until it converges to Nash-stable and obtains an approximate optimal solution. For a single coalition, the subproblem of power allocation is a concave/convex functions (DC) program \cite{5519540} that uses multivariate Taylor expansion and gradient descent to obtain approximate solutions. The phase optimization sub-problem uses a relatively acceptable local discrete phase search method.
		
		\item We compare the performance of the proposed algorithm with Non-RIS assistance and other benchmark schemes in the literature. The simulation results show that after deploying RIS to optimize the phase shift in the HCN, the interference of the system can be significantly reduced, and the communication quality can be greatly improved.
	\end{itemize}
	
	The rest of the paper is organized as follows. In Section~\ref{Sec2}, we summarize the related work of D2D communications and RIS-assisted communications. In Section~\ref{Sec3}, we build a system model, describe the channels in different frequency bands of the system, and analyze the interference involved. Then the sum rate maximization problem derived in Section~\ref{Sec4} is decomposed into mode switching, power allocation, and discrete phase optimization problems. In Section~\ref{Sec5}, the algorithm is designed for three sub-problems, and the proposed algorithm is analyzed theoretically. In Section~\ref{Sec6}, we evaluate the performance of the proposed scheme and compare it with other benchmark schemes. Finally, we conclude this paper in Section~\ref{Sec7}.	
	
	\section{Related Work}\label{Sec2}
	There have been several related works studying resource allocation and interference management for D2D communications. Ramezani \emph{et al.} \cite{ramezani2017joint} proposed an efficient power control algorithm under the constraint of the SINR requirement lower bound, and jointly optimized the power in the scenario of one cellular user and a pair of D2D users to maximize the system sum rate. Zhao \emph{et al.} \cite{zhao2015social} propose a social-aware D2D resource-sharing scheme that exploits the social network properties and centrality of communities to construct a new system design paradigm. Chen \emph{et al.} \cite{chen2018resource} proposed the problem of resource allocation between multiple pairs of D2D mm-wave and cellular bands from the perspective of game theory, and proposed a coalition formation strategy that maximizes the system sum rate in a statistical average sense. Wang \emph{et al.} \cite{wang2013position} proposed a location-based D2D communication underlying cellular system resource-sharing scheme by considering the analysis tractability of single-cell scenarios. Feng \emph{et al.} \cite{feng2017effective} consider small social communities composed of like-minded people and uses them to optimize the resource allocation of the community, arriving at a solution through bipartite graph matching and an efficient small social community resource allocation algorithm. Zhang \emph{et al.} \cite{8283645} assign sub-channels of different bandwidths to multiple D2D pairs and remote radio head users and solve the resource allocation problem with a coalition sub-game. Kazmi \emph{et al.} \cite{7890452} studied the mode selection and resource allocation of the underlying D2D network, and proposed two resource allocation algorithms based on matching theory according to the learning framework of the Markov chain. Du \emph{et al.} \cite{9849051}describe the maximum sum rate problem of UAV communication as a Nash bargaining game, and introduce Nash bargaining solution to solve the problem of resource allocation and power allocation. Based on particle swarm optimization and joint two-stage power allocation algorithm, high data rate is improved.
	
	About mm-wave communications, Deng \emph{et al.} \cite{deng2018millimeter} conduct a comprehensive study of interference characteristics and link performance in mm-wave D2D networks with varying concurrent transmission beamwidths. Chen \emph{et al.} \cite{chen2019resource} formulated the optimization problem of D2D communication spectrum resource allocation between multiple microwave frequency bands and multiple mm-wave frequency bands in HCNs. They considered the completely different propagation conditions in the two frequency bands and proposed a heuristic algorithm to maximize the system transfer rate. Compared with related work, this paper introduces a RIS-assisted system while solving the problem of resource allocation in HCNs.
	
	Some works are using RIS to assist wireless communication networks. Wu and Zhang showed in \cite{wu2019intelligent} that in RIS-assisted wireless networks, the coverage performance of the system can be significantly improved by jointly optimizing active and passive beamforming. In \cite{zhang2020reconfigurable}, the authors focused on the influence of the number of discrete phase shifts on the data rate under the assistance of RIS and studied the influence of the number of discrete phase shifts in the data rate. In \cite{wu2019towards}, the authors described the advantages of RIS based on existing technologies and propose a hardware architecture and a corresponding new signal model to implement a new RIS-assisted hybrid wireless network.
	
	In other research directions, RIS is also used to improve system performance. Hou \emph{et al.} \cite{hou2020reconfigurable} designed passive beamforming weights at the RIS, and conceives a system serving pairs of power-domain non-orthogonal multiple-access users to improve network performance.  Hou \emph{et al.} \cite{li2020reconfigurable}  investigated a joint UAV trajectory and passive beamforming design of RIS for a novel RIS-assisted UAV communication system to maximize the average achievable rate. Zappone \emph{et al.} \cite{zappone2020optimal} optimized system rate, energy efficiency, and their tradeoffs by tuning the number of RIS elements to be activated and the phase shifts they apply. Pan \emph{et al.} \cite{pan2020multicell} proposed invoking IRS at the cell boundary of multiple cells to assist downlink transmission to cell edge users while mitigating inter-cell interference, and the introduction of IRS improves cell edge user performance.
	
	However, the use of RIS for assisting D2D communication in HCNs has not been studied, and it is promising to carry out this research. Using RIS to assist D2D communication in HCNs can provide positive assistance in interference management.
	
	\section{System Model}\label{Sec3}
	\subsection{System Description}\label{Sec3-1}
	In this paper, we consider a RIS-assisted heterogeneous network in a single-cell scenario, where all users are in the cell. We focus on interference caused by users sharing the same frequency band. To mitigate interference, multiple RISs are deployed within the cell, which consists of a large number of built-in programmable elements. It reflects the signal incident on the surface and reflects the signal through a directional beam to the intended receiver. In the HCN, the network includes cellular frequency bands and mm-wave frequency bands, and D2D users can choose two frequency bands for communication. One mode is to share the spectrum with one cellular user, and the subcarrier frequency bands allocated between each cellular user are independent of each cellular user. The other is to use the mm-wave spectrum, where all D2D pairs in this mode share the same frequency band. When the D2D pairs use the cellular mode to access, the cellular user's uplink resources are shared with the D2D pairs to achieve maximum spectrum utilization. Meanwhile, it is assumed that multiple D2D pairs can simultaneously share the uplink resources of one cellular user. In the D2D communication process of the above two modes, the receiver can receive both direct signals and RIS-assisted reflected signals.
	
	\begin{figure}[!t]
		\begin{center}
			\includegraphics*[width=0.7\columnwidth,height=2.3in]{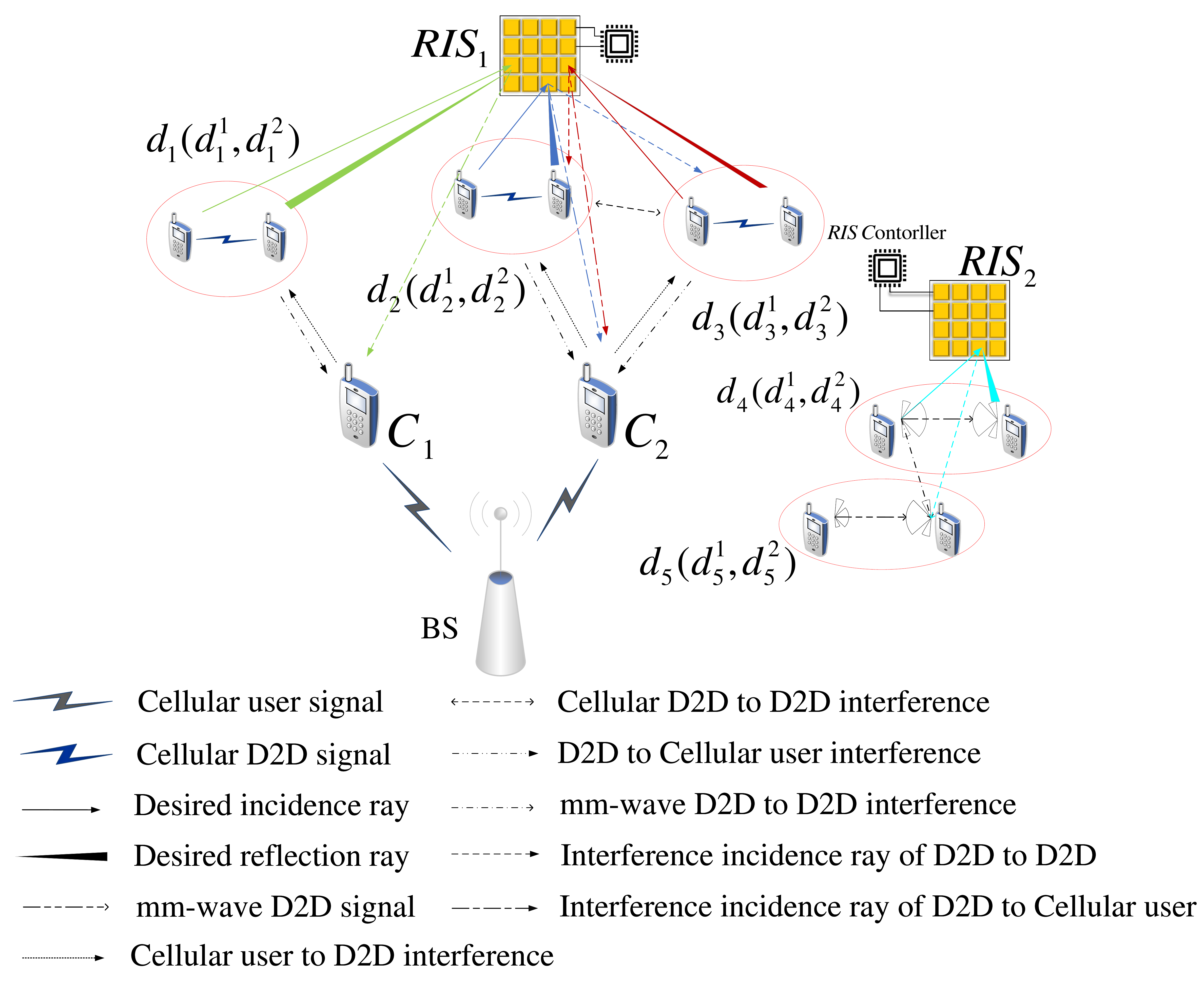}
		\end{center}
		\caption{Illustration of the resource sharing of RIS-assisted D2D communications underlaying HCN.} \label{fig1.HCN}
	\end{figure}
	
	As illustrated in Fig. \ref{HCN}, there exist two cellular users $C_1$ and $C_2$, and the D2D pair ${d_1}(d_1^1,d_1^2)$ occupies the spectrum resource of $C_1$ and uses $RIS_1$ to assist communication, while D2D pairs ${d_2}(d_2^1,d_2^2)$ and ${d_3}(d_3^1,d_3^2)$ occupy the spectrum resource of ${C_2}$ and also use $RIS_1$ to assist communication. Besides, D2D pairs ${d_4}(d_4^1,d_4^2)$ and ${d_5}(d_5^1,d_5^2)$ use the spectrum resource in the mm-wave band, assisted by $RIS_2$.
	
	In such a system, we focus on allocating uplink spectrum resources, occupied by cellular users or mm-wave band resources to D2D pairs to improve the performance of the entire network. Since D2D pairs share a spectrum with cellular users, co-channel interference is bound to occur. Corresponding co-channel interference also exists in the mm-wave frequency band. Therefore, to achieve communication goals, it is necessary to reduce some system performance to limit interference. As shown in Fig. \ref{HCN}, there are two kinds of interference in the mm-wave band, from the direct signal interference between D2D pairs and the interference through RIS reflection; in the cellular band network, there are three kinds, which are from the interference between D2D pairs occupying the uplink resources of the same cellular user, the interference of the same frequency cellular user to the D2D pair, and the interference of the D2D pair to the same frequency band cellular user.
	
	In this paper, we consider a RIS-assisted uplink single-cellular heterogeneous network, assuming that there are $C$ cellular users, denoted by ${\mathcal{C}} = \{ {C_1},{C_2},...,{C_C}\}$. Cellular users can share their uplink resources to D2D pairs, and each D2D pair can share at most uplink resources of one cellular user. Besides, we also assume that there exist $D$ D2D links, which are represented by ${\mathcal{D}} = \{ {d_1},{d_2},...,{d_D}\}$. Each D2D pair independently and randomly selects the access mode as the cellular mode or mm-wave mode. Therefore, a binary variable ${X_{c,d}}$ is defined to indicate which mode the D2D user works in, ${X_{c,d}} = 1$ indicates that the D2D pair $d$ shares the spectral resources of the cellular user $c$, otherwise ${X_{c,d}} = 0$. The links involved in the system are denoted as ${\mathcal{L}} = \{ 1,2,...,D + C - 1,D + C\}$. For communication link $i \in {\mathcal{L}}$, we denote its transmitter and receiver by ${t_i}$ and ${r_i}$, respectively. Since each D2D pair can share the uplink spectrum resources of one cellular user at most, the constraint is expressed as
	\begin{equation}
		\sum\limits_{c \in \mathcal{C}} {{X_{c,d}} \le 1,\forall d \in {\mathcal{D}}}.
		\label{eq1}
	\end{equation}
	
	We also assume that RIS is a uniform planar array composed of $N \times N$ elements, and the phase shift of each element can be tuned in real time by a set of voltage-controlled PIN diodes with ${\rm{OFF/ON}}$ states when electromagnetic waves are injected into the RIS. Depending on the voltage being regulated, the phase shift provided by the RIS will vary. The system provides $M$ RISs for assistance, denoted as ${\mathcal{M}} = \left\{ {1,2,...,M} \right\}$. For the sake of simplicity, this paper assumes that the range of adjusting the phase of each element only takes discrete values of $\left[ {{\rm{0,2}}\pi } \right]$, and has the same quantization interval between $\left[ {{\rm{0,2}}\pi } \right]$. Wherein, if the quantization bit is set to $e$ bit, there are ${2^e}$ kinds of phase switching possibilities in total. Response coefficient ${q_{{l_z},{l_y}}} = {e^{j{\theta _{{l_z},{l_y}}}}}$ at the ${l_z}{\rm{ - th}}$ row and the ${l_y}{\rm{ - th}}$ column of RIS elements where ${\theta _{{l_z},{l_y}}} = {\textstyle{{2\pi {m_{_{{l_z},{l_y}}}}} \over {{2^e} - {\rm{1}}}}}$, ${m_{{l_z},{l_y}}} = \{ 0,1,2,...,{2^e} - 1\} ,1 \le {l_z},{l_y} \le N$, and $j$ is the imaginary unit. The channels involved in communication, there are two types of channels. The first type of channel is the assisted reflection channel through the RIS. Define the channel from ${t_i}$ to ${r_i}$ that is reflected by the RIS as $g_{{l_z},{l_y}}^{{r_i},{t_i}}$. Since the channel characteristics of the mm-wave frequency band are different from those of the cellular frequency band, this paper defines the mm-wave channel from ${t_i}$ to ${r_i}$ that is reflected by RIS as $h_{{l_z},{l_y}}^{{r_i},{t_i}}$. Another type of channel is the direct channel, the mm-wave direct channel from ${t_i}$ to ${r_i}$ is ${h_{{r_i},{t_i}}}$, the cellular direct channel is ${g_{{r_i},{t_i}}}$. For the reflection channel, in order to simply represent the two reflection channels, the RIS phase and channel are combined as
	\begin{equation}
		{G_{{r_i},{t_j}}} = g_{{l_z},{l_y}}^{{r_i},{t_j}}q_{{l_z},{l_y}},1 \le i,j \le D + C,
		\label{eq2}
	\end{equation}
	\begin{equation}
		{H_{{r_i},{t_j}}} = h_{{l_z},{l_y}}^{{r_i},{t_j}}q_{{l_z},{l_y}},1 \le i,j \le D + C,
		\label{eq3}
	\end{equation}
	
	At the same time, it is assumed in this paper that the Channel State Information (CSI) is fully known, and so the above channel expressions will not affect the subsequent design of power, phase, and RIS assistance. This paper assumes a perfect understanding of CSI. This paper does not discuss the content of channel estimation. The channel estimation method is mentioned in \cite{zheng2019intelligent}. Although passive RIS elements do not have active transmitters or receivers, they can be used as active transceivers of BSs and users to estimate CSI. Other CSI estimation methods are studied in detail in \cite{wei2021channel} \cite{wei2020parallel}.
	
	\subsection{Interference Analysis}\label{Sec3-2}
	For receiver ${r_i}$ of link $i$, the received signal is the superposition of the direct signal and the reflected signal through the RIS. On the other hand, other D2D user signals and cellular user signals sharing the same cellular frequency band form superimposed interference, and all signals of other D2D users in the mm-wave band form superimposed interference. The power of the received signal $s_{{r_i}}^c$ in the cellular band D2D is expressed by accumulation as
	\begin{equation}
		\begin{array}{l}
			s_{{r_i}}^c = ({g_{{r_i},{t_i}}} + {\alpha _i}\sum\limits_{{l_z},{l_y}} {{G_{{r_i},{t_i}}}} )\sqrt {{p_i}} s_{{t_i}}^c\\
			+ \sum\limits_{d' \in {\mathcal{D}}\backslash \left\{ d \right\}} {\sum\limits_{c \in {\mathcal{C}}} {{X_{c,d'}}{X_{c,d}}} } \sum\limits_{\scriptstyle j \in {\mathcal{L}},\hfill\atop
				\scriptstyle j \ne i\hfill} {({g_{{r_i},{t_j}}} + {\alpha _j}\sum\limits_{{l_z},{l_y}} {{G_{{r_i},{t_j}}}} )} \sqrt {{p_j}} s_{{t_j}}^c\\
			{\rm{ +  }}{w_{{r_i}}},
		\end{array}
		\label{eq4}
	\end{equation}
	where $s_{{t_i}}^c$ is the unit-power transmitted symbol of cellular user from transmitter ${t_i}$, and ${p_i}$ is the transmission power of link $i$. The population can be expressed as ${\bf{P}} = {\left[ {{p_1},{p_2},...,{p_{D + C}}} \right]^T}$ is a $D{\rm{ +  }}C$ dimensional column vector, used to represent the transmission power of all links, and ${w_{{r_i}}}$ is thermal white Gaussian noise, obeying ${\cal C}{\cal N}(0,{\sigma ^2})$. Since the elements of RIS are passive, $\alpha $ represents the reflection coefficient of RIS, $\alpha \in \left[ {0,1} \right]$. Similarly, the power of the received signal of the mm-wave band D2D pair is
	\begin{equation}
		\begin{array}{l}
			s_{{r_i}}^m = ({h_{{r_i},{t_i}}} + {\alpha _i}\sum\limits_{{l_z},{l_y}} {{H_{{r_i},{t_i}}}} )\sqrt {{p_i}} s_{{t_i}}^m\\
			+ \sum\limits_{d' \in {\mathcal{D}}\backslash \left\{ d \right\}} {(1 - {X_{c,d'}})} \sum\limits_{\scriptstyle j \in {\mathcal{L}},\hfill\atop
				\scriptstyle j \ne i\hfill} {({h_{{r_i},{t_j}}} + {\alpha _j}\sum\limits_{{l_z},{l_y}} {{H_{{r_i},{t_j}}}} )} \sqrt {{p_j}} s_{{t_j}}^m\\
			+ {w_{{r_i}}},
		\end{array}
		\label{eq5}
	\end{equation}
	where $s_{{t_i}}^m$ is the unit power of the transmitted symbol in the mm-wave band${t_i}$. For cellular users, the BS received power ${s_{{r_i}}}$ is
	\begin{equation}
		\begin{array}{l}
			{s_{{r_i}}} = ({g_{{r_i},{t_i}}} + {\alpha _i}\sum\limits_{{l_z},{l_y}} {{G_{{r_i},{t_i}}}} )\sqrt {{p_i}} {s_{{t_i}}}\\
			+ \sum\limits_{d \in {\mathcal{D}}} {{X_{c,d}}} \sum\limits_{j \in {\mathcal{L}},j \ne i} {({g_{{r_i},{t_j}}} + {\alpha _j}\sum\limits_{{l_z},{l_y}} {{G_{{r_i},{t_j}}}} )} \sqrt {{p_j}} {s_{{t_j}}} + {w_{{r_i}}}.
		\end{array}
		\label{eq6}
	\end{equation}
	
	According to the above expression, the received SINR of D2D users in the cellular band can be obtained, which is expressed as (\ref{eq7}), the received  SINR of D2D users in the mm-wave band is expressed as (\ref{eq8}), and the received  SINR of cellular users is expressed as (\ref{eq9}).
	\begin{figure*}
		\begin{equation}
			\Gamma_c = \frac{{{{\left| {{g_{{r_i},{t_i}}} + {\alpha _i}\sum\limits_{{l_z},{l_y}} {{G_{{r_i},{t_i}}}} } \right|}^2}{p_i}}}{{\sum\limits_{d' \in {\mathcal{D}}\backslash \left\{ d \right\}} {\sum\limits_{c \in {\mathcal{C}}} {{X_{c,d'}}{X_{c,d}}} } \sum\limits_{\scriptstyle j \in {\mathcal{L}},\hfill\atop
						\scriptstyle j \ne i\hfill} {{{\left| {{g_{{r_i},{t_j}}} + {\alpha _j}\sum\limits_{{l_z},{l_y}} {{G_{{r_i},{t_j}}}} } \right|}^2}{p_j} + {\sigma ^2}} }},
			\label{eq7}
		\end{equation}
		\begin{equation}
			\Gamma_m = \frac{{{{\left| {{h_{{r_i},{t_i}}} + {\alpha _i}\sum\limits_{{l_z},{l_y}} {{H_{{r_i},{t_i}}}} } \right|}^2}{p_i}}}{{\sum\limits_{d' \in {\mathcal{D}}\backslash \left\{ d \right\}} {(1 - {X_{c,d'}})} \sum\limits_{\scriptstyle j \in {\mathcal{L}},\hfill\atop
						\scriptstyle j \ne i\hfill} {{{\left| {{h_{{r_i},{t_j}}} + {\alpha _j}\sum\limits_{{l_z},{l_y}} {{H_{{r_i},{t_j}}}} } \right|}^2}{p_j} + {\sigma ^2}} }},
			\label{eq8}
		\end{equation}
		\begin{equation}
			\Gamma_{cu} = \frac{{{{\left| {{g_{{r_i},{t_i}}} + {\alpha _i}\sum\limits_{{l_z},{l_y}} {{G_{{r_i},{t_i}}}} } \right|}^2}{p_i}}}{{\sum\limits_{d \in {\mathcal{D}}} {{X_{c,d}}} \sum\limits_{j \in {\mathcal{L}},j \ne i} {{{\left| {{g_{{r_i},{t_j}}} + {\alpha _j}\sum\limits_{{l_z},{l_y}} {{G_{{r_i},{t_j}}}} } \right|}^2}{p_j} + {\sigma ^2}} }}.
			\label{eq9}
		\end{equation}
	\end{figure*}
	
	According to the Shannon capacity formula, the possible channel capacity of D2D users in the cellular band $R_{{r_i}}^{dc}$ can be written, expressed as
	\begin{equation}
		R_{{r_i}}^{dc} = {W_c}{\log _2}\left( {1 + \Gamma_c} \right).
		\label{eq10}
	\end{equation}
	The possible channel capacity of D2D users in the mm-wave band is expressed as
	\begin{equation}
		R_{{r_i}}^{dm} = {W_m}{\log _2}\left( {1 + \Gamma_m} \right).
		\label{eq11}
	\end{equation}
	The possible channel capacity of cellular users is expressed as
	\begin{equation}
		R_{{r_i}}^c = {W_c}{\log _2}\left( {1 + \Gamma_{cu}} \right).
		\label{eq12}
	\end{equation}
	
	The total sum rate of the system can be expressed as (\ref{eq13}), where ${P_{out:i,j}}$ represents the connection interruption probability of the D2D user between the transmitter and receiver in the mm-wave band. Here ${P_{out:i,j}} = 1 - {e^{ - {\beta _1}{l_{ij}}}}$, where ${l_{ij}}$ is the distance between users $i$ to $j$, and ${\beta _1}$ is the parameter used to reflect the density and size of obstacles, which cause blockages and cause interruptions \cite{jung2016connectivity}.
	\begin{figure*}[!t]
		\begin{equation}
			R = \sum\limits_{i{\rm{ = 1}}}^{D + C} {\left( {\left( {1 - {A_i}} \right)R_{{r_i}}^c + {A_i}\left( {{X_{c,d}}R_{{r_i}}^{dc} + (1 - {X_{c,d}})(1 - {P_{out:d,d}})R_{{r_i}}^{dm}} \right)} \right)}
			\label{eq13}
		\end{equation}
	\end{figure*}
	
	To sum up, the interference involved in the whole system is mainly the interference between D2D users in the same frequency band, the interference between cellular users and D2D users in the same frequency band, and the interference only involves the interference of D2D users in the same frequency band in the mm-wave band.
	
	\section{Problem Formulation and Decomposition}\label{Sec4}	
	\subsection{Sum Rate Maximization Problem Formulation}\label{Sec4-1}
	Based on the problem planning of the above system model, the optimization goal is to maximize the system sum rate. Obviously, the sum rate of the system is related to ${X_{c,d}}$. At the same time, through RIS assistance and power control, the variables in the problem also include transmitting power ${p_i}$ and conversion phase ${\theta _{{l_z},{l_y}}}$. In this paper, all phase shift values are written as a set ${\Theta } = \left\{ {{\theta _{{l_z},{l_y}}},1 \le {l_z},{l_y} \le N} \right\}$. The sum rate maximization problem can be written as the equation (\ref{eq14}).

	Constraint (a) indicates the minimum SINR constraint for cellular users and D2D users under the condition of satisfying QoS, and the limited maximum power is expressed in constraint (b). In constraint (c), the phase is a discrete variable, and constraint (d) is the working mode constraint of the D2D user, which is a binary variable. Since the SINR and transmission power shows continuous properties, phase, and ${X_{c,d}}$ are discrete constraints, the whole problem is a mixed integer nonlinear optimization problem. The variables ${\bf{P}}$ and ${\Theta }$ are coupled, which is difficult to solve.
	\begin{align}\label{eq14}
	&\mathop {\max }\limits_{{\bf{P}},{X_{c,d}},{{\Theta }}} R \nonumber \\
	& s.t. \nonumber \\
	& (a)~\{\Gamma_{c},\Gamma_{m},\Gamma_{cu}\} \ge {\gamma _{min}}, \nonumber \\
	& (b)~0 \le {p_i} \le {P_{\max }},{\rm{ }}\forall i = 1,2,...,D + C ,\nonumber \\
	& (c)~\sum\limits_{c \in \mathcal{C}} {{X_{c,d}} \le 1,{\rm{ }}\forall d \in {\mathcal{D}}} ;{X_{c,d}} \in \left\{ {0,1} \right\},{\rm{ }}\forall d \in {\mathcal{D}}, \nonumber \\
	& (d)~{q_{{l_z},{l_y}}} = {e^{j{\theta _{{l_z},{l_y}}}}},{\theta _{{l_z},{l_y}}} = \frac{{2\pi {m_{{l_z},{l_y}}}}}{{{2^e} - 1}}, \nonumber\\
	&{\kern 1pt} {\kern 1pt} {\kern 1pt} {m_{{l_z},{l_y}}} = \{ 0,1,2,...,{2^e} - 1\} ,1 \le {l_z},{l_y} \le N.
	\end{align}
	\subsection{Problem Decomposition}\label{Sec4-2}
	In order to solve the proposed mixed integer nonlinear programming problem more efficiently, this paper deconstructs the problem into three solvable sub-problems.
	
	\subsubsection{Mode Switch}\label{Sec4-2-1}
	This sub-problem is used to solve the working mode switching problem of D2D users in HCNs under SINR constraints. By switching D2D users to communicate in the cellular frequency band or the mm-wave frequency band, the system sum rate is maximized. When switching modes, the transmit power of the D2D user and the RIS-assisted phase are fixed, to determine the working mode of the D2D user of the entire system, and then the problem is transformed into
	\begin{equation}\label{eq15}
		\begin{array}{l}
			\mathop {\max }\limits_{{X_{c,d}}} R\\
			s.t.{\rm{   }}\\
			\left( a \right)\{\Gamma_{c},\Gamma_{m},\Gamma_{cu}\} \ge {\gamma _{min}}\\
			\left( b \right)\sum\limits_{c \in C} {{X_{c,d}} \le 1,{\rm{ }}\forall d \in {\mathcal{D}}} ;{X_{c,d}} \in \left\{ {0,1} \right\},{\rm{ }}\forall d \in {\mathcal{D}}.
		\end{array}
	\end{equation}
	The above problem is still a mixed integer nonlinear programming problem, but after the constraints are reduced, the suboptimal solution can be approximated to the optimal solution by methods such as game theory, and the game theory model is introduced to solve it comprehensively.
	\subsubsection{Power Allocation}\label{Sec4-2-2}
	This sub-problem is used to allocate the transmission power of users, and solve the problem of system sum rate maximization under the condition of satisfying the SINR constraint and the maximum power constraint. In this sub-problem, the phase and the user's working mode are fixed and the problem is reduced to
	\begin{equation}\label{eq16}
		\begin{array}{l}
			\mathop {\max }\limits_{\bf{P}} R\\
			s.t.{\rm{   }}\\
			\left( a \right)\{\Gamma_{c},\Gamma_{m},\Gamma_{cu}\} \ge {\gamma _{min}}\\
			\left( b \right)0 \le {p_i} \le {P_{\max }},{\rm{ }}\forall i = 1,2,...,D + C.
		\end{array}
	\end{equation}
	After simplification, the sub-problem constraints only involve continuous quantities, and the objective function is also continuous, but the whole problem is still non-convex, and it is not easy to solve. Generally, it is necessary to try mathematical transformation to convert it into convex optimization or approximate problem-solving. The specific problem-solving method will be discussed in detail later.
	\subsubsection{Discrete Phase Shift Optimization}\label{Sec4-2-3}
	By fixing the working mode and transmit power of the D2D user, adjusting the phase of each element of the RIS provides $2^e$ variations. So far, the optimization problem is transformed into
	\begin{equation}\label{eq17}
		\begin{array}{l}
			\mathop {\max }\limits_\Theta  R\\
			s.t.{\rm{   }}\\
			\left( a \right)\{\Gamma_{c},\Gamma_{m},\Gamma_{cu}\} \ge {\gamma _{min}}\\
			\left( b \right){q_{{l_z},{l_y}}} = {e^{j{\theta _{{l_z},{l_y}}}}},{\theta _{{l_z},{l_y}}} = \frac{{2\pi {m_{{l_z},{l_y}}}}}{{{2^e} - 1}},\\
			{\rm{ }}{m_{{l_z},{l_y}}} = \{ 0,1,2,...,{2^e} - 1\} ,1 \le {l_z},{l_y} \le N.
		\end{array}
	\end{equation}
	Considering the complexity of the problem, this paper adopts the local search method to solve the sub-problem, which will be discussed in the following sections.
	
	\section{Sum Rate Maximization Approach}\label{Sec5}
	In this section, the solutions to the three sub-problems proposed above will be discussed in detail. The mode switch problem will be solved in Subsection \ref{Sec5-1}. The power allocation problem will be solved in Subsection \ref{Sec5-2}. The discrete phase shift problem will be solved in Subsection \ref{Sec5-3}. A sum rate maximization algorithm in an alternate iterative method will be proposed until the algorithm converges and a suboptimal solution is obtained in Subsection \ref{Sec5-4}. As shown in Fig. \ref{block diagram}, the block diagram shows the basic relationship of each sub-algorithm proposed in this section. 	
	\begin{figure}[!t]
		\begin{center}
			\includegraphics*[width=0.9\columnwidth,height=2.3in]{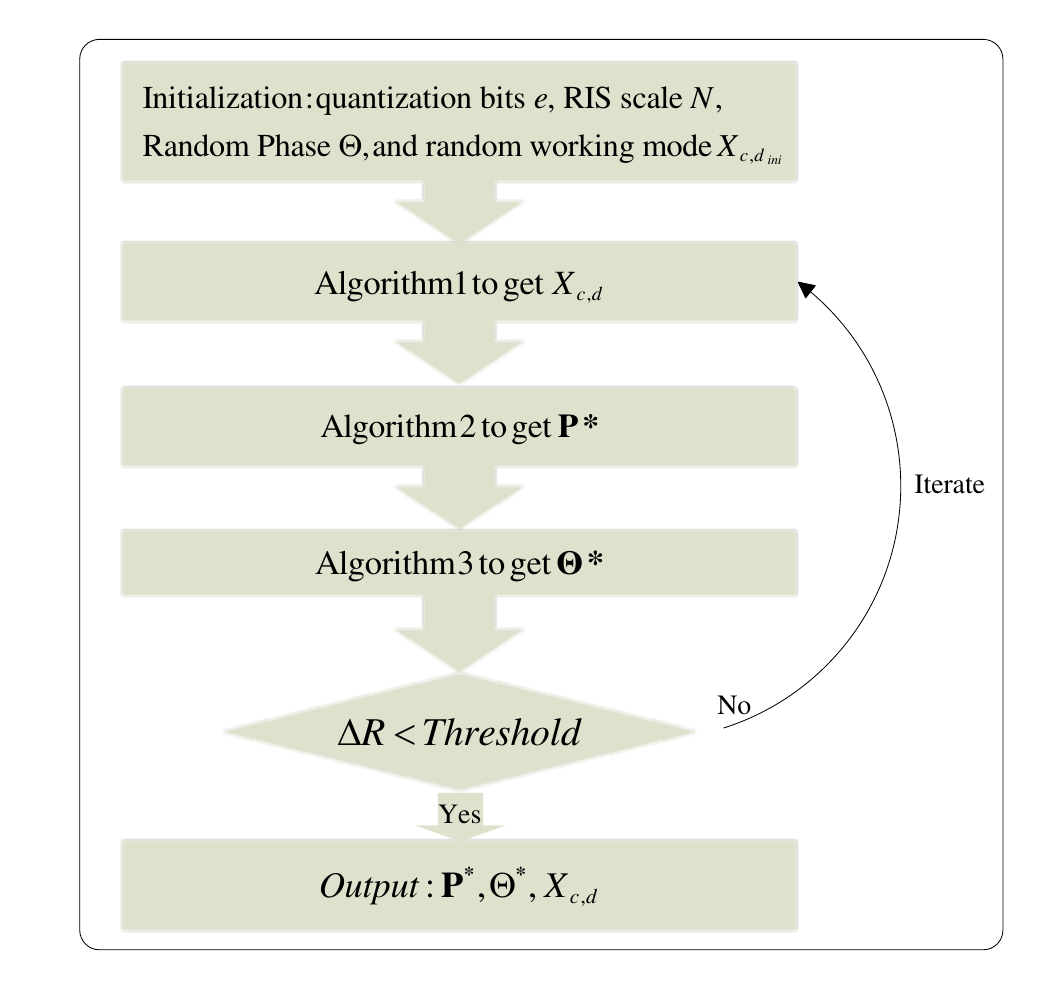}
		\end{center}
		\caption{A block diagram of the proposed algorithm.}
		\label{fig2.BlockDiagram}
	\end{figure}	

	\subsection{Coalition Game Formation Algorithm}\label{Sec5-1}
	From the perspective of game theory, this subsection uses the coalitional game to solve resource sharing, i.e., the switching of the working mode of D2D users. Based on this, a coalition formation algorithm is proposed.
	
	The whole optimization problem is aimed at maximizing the sum rate of the overall system and based on this condition, a coalition game theory model is introduced, in which D2D pairs tend to improve system performance in a coalition-forming manner. In the system studied in this paper, there are $C$ cellular users and $D$ D2D users. D2D users can choose to occupy the resources of any cellular user or use the mm-wave band resources. Therefore, it is assumed that there are $C+1$ coalitions formed by D2D users $F = \{ {F_{{c_1}}},{F_{{c_2}}},...,{F_{{c_C}}},{F_{{c_{C + 1}}}}\} $. There are no overlapping D2D users in any two coalitions, that is, any user exists in only one coalition $\bigcup\nolimits_{x = 1}^{C + 1} {{F_{{c_x}}}}  = {\mathcal{D}}$. Divide the coalitions into two groups, the first group consists of a coalition ${F_c}$, ${F_c} \subset F(c \in {\mathcal{C}})$ to share spectrum resources with cellular users. The uplink transmission rate of cellular user $c$ can be written as the equation (\ref{eq18}). The uplink transmission rate of D2D pair $d$ can be expressed as the equation (\ref{eq19}).
	\begin{figure*}[!t]
		\begin{equation}
			{R_c} = {W_c}{\log _2}\left( {1 + \frac{{{{\left| {{g_{{r_i},{t_i}}} + {\alpha _i}\sum\limits_{{l_z},{l_y}} {{G_{{r_i},{t_i}}}} } \right|}^2}{p_i}}}{{\sum\limits_{d \in {\mathcal{D}}} {{X_{c,d}}} \sum\limits_{j \in {\mathcal{L}},j \ne i} {{{\left| {{g_{{r_i},{t_j}}} + {\alpha _j}\sum\limits_{{l_z},{l_y}} {{G_{{r_i},{t_j}}}} } \right|}^2}{p_j} + {\sigma ^2}} }}} \right)
			\label{eq18}
		\end{equation}
		\begin{equation}
			\begin{array}{l}
				{R_d} = {X_{c,d}}{W_c}{\log _2}\left( {1 + \frac{{{{\left| {{g_{{r_i},{t_i}}} + {\alpha _i}\sum\limits_{{l_z},{l_y}} {{G_{{r_i},{t_i}}}} } \right|}^2}{p_i}}}{{\sum\limits_{d' \in {\mathcal{D}}\backslash \left\{ d \right\}} {\sum\limits_{c \in {\mathcal{C}}} {{X_{c,d'}}{X_{c,d}}} } \sum\limits_{j \in {\mathcal{L}},j \ne i} {{{\left| {{g_{{r_i},{t_j}}} + {\alpha _j}\sum\limits_{{l_z},{l_y}} {{G_{{r_i},{t_j}}}} } \right|}^2}{p_j} + {\sigma ^2}} }}} \right)\\
				+ (1 - {X_{c,d}})(1 - {P_{out:d,d}}){W_m}{\log _2}\left( {1 + \frac{{{{\left| {{h_{{r_i},{t_i}}} + {\alpha _i}\sum\limits_{{l_z},{l_y}} {{H_{{r_i},{t_i}}}} } \right|}^2}{p_i}}}{{\sum\limits_{d' \in {\mathcal{D}}\backslash \left\{ d \right\}} {(1 - {X_{c,d'}})} \sum\limits_{j \in {\mathcal{L}},j \ne i} {{{\left| {{h_{{r_i},{t_j}}} + {\alpha _j}\sum\limits_{{l_z},{l_y}} {{H_{{r_i},{t_j}}}} } \right|}^2}{p_j} + {\sigma ^2}} }}} \right)
			\end{array}
			\label{eq19}
		\end{equation}
	\end{figure*}
	In the reflection channel, the RIS reflection coefficient in the model is set to 1 in the cellular band and 0.8 in the mm-wave band.
	
	From (\ref{eq18}) and (\ref{eq19}), it can be obtained that as the number of D2D users in a single coalition increases, the interference of users in the coalition also increases. In the coalition game, if all D2D pairs form a big coalition and share the uplink resources of the same cellular user, and then due to the increase in interference, no D2D user can make a greater contribution to the system. Therefore, the D2D pair has no incentive to form a major coalition. The mm-wave band communication rate is 2 orders of magnitude higher than the cellular band, so more D2D pairs will choose to communicate in the mm-wave band, which may lead to empty coalitions in some cellular bands. Based on this, a coalition game with transferable utility is established. As a participant in the game, D2D users choose different working bands to maximize the system and speed. Therefore, we define the coalition game with transferable utility as follows.
	\begin{definition}	\label{def1}
		The concept of coalition games with transferable utility was first proposed by Von Neumann and Morgenstern \cite{von2007theory}. Resource allocation for D2D users in HCNs is constructed by a coalition game pair with transferable utility $({\mathcal{D}},R)$ where ${\mathcal{D}}$ is the coalition user and $R$ is the win function. The game theory consists of these two elements. $A$ is a value that exists exactly and represents the interests of the entire coalition and the interests of the entire coalition can be distributed to each member in any random way. The coalition game defined by the resource-sharing relationship will be proposed next.
	\end{definition}
	\begin{definition}	\label{def2}
		Coalition game defined by resource sharing relationship: A coalition game with transferable utility in resource allocation for D2D users is defined by $({\mathcal{D}},R,F)$, where $R$ is the transferable utility and $F$ is the coalition distribution $F = \{ {F_{{c_1}}},{F_{{c_2}}},...,{F_{{c_C}}},{F_{{c_{C + 1}}}}\} $ and for any $x \ne x'$ there are ${F_{{c_x}}} \cap {F_{{c_{x'}}}} = \emptyset $ and $\bigcup\nolimits_{x = 1}^{C + 1} {{F_{{c_x}}}} = {\mathcal{D}}$. Satisfying the condition of the minimum SINR while allocating resources means meeting the condition of the minimum transmission rate when the user bandwidth is fixed. This is a shared resource strategy decided by each D2D user based on system and utility.
	\end{definition}
	\begin{definition}	\label{def3}
		For any pair of D2D users, the preference relation ${ \succ _i}$ is defined as a complete, transitive, and reflexive binary relation across the entire formable coalition. D2D users have the right to choose to join or leave a coalition according to their preferences, in other words, D2D users are more inclined to become a member of the coalition they prefer. Therefore, $i \in {\mathcal{D}}, {F_c}{ \succ _i}{F_{c'}}$ represents that D2D pair $i$ are more inclined to join coalition ${F_c}$ than coalition ${F_{c'}}$. The considered situation does not include the same preference of the D2D users for the two coalitions. In this case, the D2D users are more conservative and tend to the current coalition. The preference for switching is expressed as follows
		\begin{equation}\label{eq20}
			{F_c}{ \succ _i}{F_{c'}} \Leftrightarrow R({F_c}) + R({F_{c'}}\backslash i) > R({F_{c'}}) + R({F_c}\backslash i).
		\end{equation}
	\end{definition}
	
	In order to form the final coalition, the definition of the handover operation is made.
	\begin{definition}	\label{def4}
		For the D2D pairs set ${\mathcal{D}}$, the initial distribution $F = \{ {F_{{c_1}}},{F_{{c_2}}},...,{F_{{c_C}}},{F_{{c_{C + 1}}}}\} $ is given. If D2D users $i \in {\mathcal{D}}$ express the idea of switching from coalition ${F_c}$ to coalition ${F_{c'}}$, and ${F_c} \ne {F_{c'}}$ , then the current total distribution $F$ is modified to the new distribution $F'$, $F' = (F\backslash \{ {F_{c'}},{F_c}\} ) \cup \{ {F_c}\backslash \{ i\} ,{F_{c'}} \cup \{ i\} \} $.
	\end{definition}
	
	Initialize the system, randomly assign the coalition of any D2D user $F = \{ {F_{{c_1}}},{F_{{c_2}}},...,{F_{{c_C}}},{F_{{c_{C + 1}}}}\} $, and assume that the D2D user $i \in {\mathcal{D}}$ is in the coalition ${F_c}$. Then randomly select other coalitions ${F_{c'}}$, and assume that their preference relationship is ${F_{c'}}{ \succ _i}{F_c}$, ${F_{c'}} \subset F$, ${F_c} \ne {F_{c'}}$ means that user $i$ performs the operation in Definition \ref{def4} and updates the new coalition distribution. In practice, the switch is performed only when $R({F_{c'}}) + R({F_c}\backslash i) > R({F_c}) + R({F_{c'}}\backslash i)$ is satisfied. In such a mechanism, each D2D pair can leave the current coalition and join another coalition, and the sum rate of the new coalition is strictly greater than that of the old coalition, to ensure a greater contribution to the entire system. The purpose is to find a fixed coalition structure that maximizes the interests of the overall system rather than individual interests. The entire coalition game formation algorithm is summarized as Algorithm \ref{alg1}.
	
	\begin{algorithm}[t]
		\caption{Coalition Game Formation Algorithm}
		\label{alg1}
		\begin{algorithmic}[1]
			\STATE Initialization, given the initial coalition ${F_{ini}}$ of all D2D users ${\mathcal{D}}$;
			\STATE Assign the current distribution ${F_{ini}}$ to ${F_{cur}}$, $num = 0$;
			\REPEAT
			\STATE  Select D2D pair $i \in {\mathcal{D}}$ in a predetermined order, ${F_c} \subset {F_{cur}}$;
			\STATE  Randomly search for another possible coalition ${F_{c'}} \ne {F_c},{F_{c'}} \subset {F_{cur}}$ ;
			\STATE  Caculate $R({F_c})$ and $R({F_{c'}})$;
			\IF {The switch operation form ${F_c}$ to ${F_{c'}}$ satisfying ${F_{c'}}{ \succ _i}{F_c}$}
			\STATE  $num = 0$;
			\STATE  D2D pair $i$ leaves current coalition ${F_c}$ and joins coalition ${F_{c'}}$;
			\STATE  Update current user distribution $({F_{cur}}\backslash \{ {F_{c'}},{F_c}\} ) \cup \{ {F_c}\backslash \{ i\} ,{F_{c'}} \cup \{ i\} \}  \to {F_{cur}}$;
			\ELSE
			\STATE  $num = num + 1$;
			\ENDIF
			\UNTIL {The existing user distribution satisfies the Nash-stable}.
		\end{algorithmic}
	\end{algorithm}
	
	In Algorithm \ref{alg1}, first, the initial coalition ${F_{ini}}$ is randomly assigned, then in step 4, a pair of D2D users are selected in a predetermined order to save the current coalition ${F_c}$, and then another possible coalition ${F_{c'}}$ is randomly selected in step 5. Then obtain the channel information of ${F_c}$ and ${F_{c'}}$ from the BS, calculate the old and new coalition sum rates, and decide whether to switch to the new coalition. If the conditions are met, update the current distribution of users in the coalition, and reset the count of consecutive unsuccessful operations to zero. Otherwise, add 1 to the count of consecutive unsuccessful operations. When $num$ is equal to ten times the number of D2D users, the algorithm stops iterating, jumps out of the loop, and finally reaches Nash-stable within a limited number of exchanges to obtain a stable distribution.
	
	\subsection{Single Coalition Power Allocation Algorithm}\label{Sec5-2}
	In the power allocation subproblem, it is simplified because the discrete variables phase and mode of operation are removed, but the subproblem is still nonconvex concerning ${\bf{P}}$. For each coalition, regardless of mm-wave or cellular users, a simple mathematical transformation of the problem is based on experience. In the objective function, both the cellular transmission rate and the D2D transmission rate can be written as the difference between two concave functions. For the cellular band and the mm-wave band coalition, due to the different channel expressions, the problem formulation is different, but the representation of this sub-problem is similar in form. Assuming that there are $K$ users in the current coalition, the power allocation problem for the cellular frequency band is transformed as (\ref{eq21}). In the mm-wave band coalition, the power allocation problem can be written as (\ref{eq22}).
	\begin{figure*}
		\begin{equation}\label{eq21}
			\begin{array}{l}
				\mathop {\max }\limits_{\bf{P}} \sum\limits_{i = 1}^{K + 1} {{R_i}} \\
				= \mathop {\max }\limits_{\bf{P}} \sum\limits_{i = 1}^{K + 1} {{{\log }_2}\left( {1 + \frac{{{{\left| {{g_{{r_i},{t_i}}} + {\alpha _i}\sum\limits_{{l_z},{l_y}} {{G_{{r_i},{t_i}}}} } \right|}^2}{p_i}}}{{\sum\limits_{j \in {F_c} \cup {C_i},j \ne i} {{{\left| {{g_{{r_i},{t_j}}} + {\alpha _j}\sum\limits_{{l_z},{l_y}} {{G_{{r_i},{t_j}}}} } \right|}^2}{p_j} + {\sigma ^2}} }}} \right)} \\
				=  - \mathop {\min }\limits_{\bf{P}} \sum\limits_{i = 1}^{K + 1} {\left[ {\lg {\kern 1pt} {\kern 1pt} {\kern 1pt} \left( {\sum\limits_{j \in {F_c} \cup {C_i},j \ne i} {{{\left| {{g_{{r_i},{t_j}}} + {\alpha _j}\sum\limits_{{l_z},{l_y}} {{G_{{r_i},{t_j}}}} } \right|}^2}{p_j} + {\sigma ^2}} } \right)} \right.} \\
				{\kern 1pt} {\kern 1pt} {\kern 1pt} {\kern 1pt} {\kern 1pt} {\kern 1pt} {\kern 1pt} {\kern 1pt} {\kern 1pt} \left. { - \lg {\kern 1pt} {\kern 1pt} \left( {{{\left| {{g_{{r_i},{t_i}}} + {\alpha _i}\sum\limits_{{l_z},{l_y}} {{G_{{r_i},{t_i}}}} } \right|}^2}{p_i} + \sum\limits_{j \in {F_c} \cup {C_i},j \ne i} {{{\left| {{g_{{r_i},{t_j}}} + {\alpha _j}\sum\limits_{{l_z},{l_y}} {{G_{{r_i},{t_j}}}} } \right|}^2}{p_j} + {\sigma ^2}} } \right)} \right]
			\end{array}
		\end{equation}
		\begin{equation}\label{eq22}
			\begin{array}{l}
				\mathop {\max }\limits_{\bf{P}} \sum\limits_{i = 1}^K {{R_i}} \\
				= \mathop {\max }\limits_{\bf{P}} \sum\limits_{i = 1}^K {{{\log }_2}\left( {1 + \frac{{{{\left| {{h_{{r_i},{t_i}}} + {\alpha _i}\sum\limits_{{l_z},{l_y}} {{H_{{r_i},{t_i}}}} } \right|}^2}{p_i}}}{{\sum\limits_{j \in {F_c},j \ne i} {{{\left| {{h_{{r_i},{t_j}}} + {\alpha _j}\sum\limits_{{l_z},{l_y}} {{H_{{r_i},{t_j}}}} } \right|}^2}{p_j} + {\sigma ^2}} }}} \right)} \\
				=  - \mathop {\min }\limits_{\bf{P}} \sum\limits_{i = 1}^K {\left[ {\lg {\kern 1pt} {\kern 1pt} {\kern 1pt} \left( {\sum\limits_{j \in {F_c},j \ne i} {{{\left| {{h_{{r_i},{t_j}}} + {\alpha _j}\sum\limits_{{l_z},{l_y}} {{H_{{r_i},{t_j}}}} } \right|}^2}{p_j} + {\sigma ^2}} } \right)} \right.} \\
				{\kern 1pt} {\kern 1pt} {\kern 1pt} {\kern 1pt} {\kern 1pt} {\kern 1pt} {\kern 1pt} {\kern 1pt} {\kern 1pt} \left. { - \lg {\kern 1pt} {\kern 1pt} \left( {{{\left| {{h_{{r_i},{t_i}}} + {\alpha _i}\sum\limits_{{l_z},{l_y}} {{H_{{r_i},{t_i}}}} } \right|}^2}{p_i} + \sum\limits_{j \in {F_c},j \ne i} {{{\left| {{h_{{r_i},{t_j}}} + {\alpha _j}\sum\limits_{{l_z},{l_y}} {{H_{{r_i},{t_j}}}} } \right|}^2}{p_j} + {\sigma ^2}} } \right)} \right]
			\end{array}
		\end{equation}
	\end{figure*}
	
	Based on the above mathematical transformation, for the sake of simplicity, only the cellular band is selected as an example, and two functions ${g_i}({\bf{P}})$ and ${\varphi _i}({\bf{P}})$ are defined for this purpose.
	\begin{equation}\label{eq23}
		{g_i}({\bf{P}}) = \lg \left( {\sum\limits_{j \in {F_c} \cup {C_i},j \ne i} {{{\left| {{g_{{r_i},{t_j}}} + {\alpha _i}\sum\limits_{{l_z},{l_y}} {{G_{{r_i},{t_j}}}} } \right|}^2}{p_j} + {\sigma ^2}} } \right)
	\end{equation}
	\begin{equation}\label{eq24}
		\begin{array}{l}
			{\varphi _i}({\bf{P}}) = \lg {\kern 1pt} {\kern 1pt} \left( {{{\left| {{g_{{r_i},{t_i}}} + {\alpha _i}\sum\limits_{{l_z},{l_y}} {{G_{{r_i},{t_i}}}} } \right|}^2}{p_i}} \right.\\
			\left. { + \sum\limits_{j \in {F_c} \cup {C_i},j \ne i} {{{\left| {{g_{{r_i},{t_j}}} + {\alpha _j}\sum\limits_{{l_z},{l_y}} {{G_{{r_i},{t_j}}}} } \right|}^2}{p_j} + {\sigma ^2}} } \right)
		\end{array}
	\end{equation}
	So far, problem (\ref{eq16}) can be written as (\ref{eq25}), thus the problem is transformed into a DC planning problem, which is still difficult to solve.
	\begin{equation}\label{eq25}
		\begin{array}{l}
			\mathop {\min }\limits_{\bf{P}} \sum\limits_{i = 1}^{{F_{{c_i}}} + 1} {{f_i}({\bf{P}}) = {g_i}({\bf{P}}) - {\varphi _i}({\bf{P}})} \\
			s.t.\\
			{\kern 1pt} (a)~{g_i}({\bf{P}}) - {\varphi _i}({\bf{P}}) \le  - \lg ({\gamma _{min}} + 1),\forall i = 1,2,...,K,K + 1\\
			{\kern 1pt} (b)~0 \le {p_i} \le {P_{max}},\forall i = 1,2,...,K,K + 1
		\end{array}
	\end{equation}
	Therefore, the first-order Taylor expansion is used to approximate it as a convex function, which provides an upper bound for the objective function, and then the optimal solution is continuously approached from the upper bound. Under this approximation, both the objective function and the constraints are convex concerning the variable $\bf{P}$. ${g_i}({\bf{P}})$ can be written as (\ref{eq26}) after $n$ iterations, and $f_i^{(n)}({\bf{P}})$ can be expressed as (\ref{eq27}).
	\begin{equation}\label{eq26}
		{g_i}({\bf{P}}) = {g_i}({{\bf{P}}^{(n)}}) + {\left. {\sum\limits_{i = 1}^{K + 1} {\frac{{\partial {g_i}({\bf{P}})}}{{\partial {p_i}}}} } \right|_{{\bf{P}} = {{\bf{P}}^{(n)}}}}({p_i} - p_i^{(n)})
	\end{equation}
	\begin{equation}\label{eq27}
		f_i^{(n)}({\bf{P}}) = {g_i}({{\bf{P}}^{(n)}}) + {\left. {\sum\limits_{i = 1}^{K + 1} {\frac{{\partial {g_i}({\bf{P}})}}{{\partial {p_i}}}} } \right|_{{\bf{P}} = {{\bf{P}}^{(n)}}}}({p_i} - p_i^{(n)}) - {\varphi _i}({\bf{P}})
	\end{equation}
	According to (\ref{eq26}) and (\ref{eq27}), the problem can be further written as
	\begin{equation}\label{eq28}
		\begin{array}{l}
			\mathop {\min }\limits_{\bf{P}} \sum\limits_{i = 1}^{K + 1} {f_i^{(n)}({\bf{P}})} \\
			s.t.{\kern 1pt} {\kern 1pt} {\kern 1pt} {\kern 1pt} {\kern 1pt} \\
			{\kern 1pt} {\kern 1pt} (a)~f_i^{(n)}({\bf{P}}) \le  - \lg ({\gamma _{min}} + 1),\forall i = 1,2,...,K,K + 1\\
			{\kern 1pt} {\kern 1pt} (b)~0 \le {p_i} \le {P_{max}},\forall i = 1,2,...,K,K + 1.
		\end{array}
	\end{equation}
	
	So far, the power distribution subproblem has been transformed into a convex optimization problem that can be solved by the Lagrangian method. By constructing a Lagrangian unconstrained function and then using the iterative gradient descent method, since the objective function is a decreasing function, it can converge to a static point in the feasible region of ${\bf{P}}$ after $n$ iterations. The Lagrangian unconstrained function is defined as follows
	\begin{equation}\label{eq29}
		{L^{(n)}}({\bf{P}},{{\bf{\lambda }}^{\left( n \right)}}) = \sum\limits_{i = 1}^{K + 1} {f_i^{(n)}({\bf{P}}){\rm{ + }}\lambda _i^{(n)}\left[ {f_i^{(n)}({\bf{P}}) + \lg \left( {{\gamma _{min}} + 1} \right)} \right]},
	\end{equation}
	where $\lambda _i^{(n)}, i = 1,2,...,K,K + 1$ refers to the Lagrange multiplier corresponding to constraint $a$ in the (\ref{eq28}). By bringing the transmission power ${{\bf{P}}^{(n + 1)}}$ into (\ref{eq29}), the Lagrangian function of the Langrangian multiplier is as follow
	\begin{equation}\label{eq30}
		\begin{array}{l}
			{L^{(n)}}({{\bf{P}}^{(n + 1)}},{\bf{\lambda }})\\
			= \sum\limits_{i = 1}^{K + 1} {\left[ {{g_i}({{\bf{P}}^{(n)}}) + {{\left. {\sum\limits_{i = 1}^{K + 1} {\frac{{\partial {g_i}({\bf{P}})}}{{\partial {p_i}}}} } \right|}_{{\bf{P}} = {{\bf{P}}^{(n)}}}}({p_i} - p_i^{(n)})} \right.} \\
			\left. {\begin{array}{*{20}{c}}
					{}\\
					{}
				\end{array} - {\varphi _i}({{\bf{P}}^{(n + 1)}})} \right]\\
			\begin{array}{*{20}{c}}
				{}\\
				{}
			\end{array}{\kern 1pt} {\kern 1pt} {\kern 1pt} {\rm{ + }}\sum\limits_{i = 1}^{K + 1} {\lambda _i^{(n)}\left[ {{g_i}({{\bf{P}}^{(n)}}) + {{\left. {\sum\limits_{i = 1}^{K + 1} {\frac{{\partial {g_i}({\bf{P}})}}{{\partial {p_i}}}} } \right|}_{{\bf{P}} = {{\bf{P}}^{(n)}}}}({p_i} - p_i^{(n)})} \right.} \\
			\left. {\begin{array}{*{20}{c}}
					{}\\
					{}
				\end{array} - {\varphi _i}({{\bf{P}}^{(n + 1)}}) + \lg \left( {{\gamma _{min}} + 1} \right)} \right].
		\end{array}
	\end{equation}
	The details of solving this DC problem will be presented in Algorithm \ref{alg2}.
	
	\begin{algorithm}[t]
		\caption{Single Coalition Power Allocation Algorithm}
		\renewcommand{\algorithmicrequire}{\textbf{Initialization:}}
		\renewcommand{\algorithmicensure}{\textbf{Output:}}
		\label{alg2}
		\begin{algorithmic}[1]
			\REQUIRE
			$n=0, {\mu ^{(0)}}=100, {\delta ^{(0)}}=50, \lambda _i^{(0)}=100, \forall i=1,...,K,K + 1$;
			\ENSURE
			${{\bf{P}}^{\bf{*}}}$
			\STATE  Transform a complex primal problem into a DC problem ;
			\STATE  Defining Lagrangian Unconstrained Functions ${L^{(n)}}({\bf{P}},{{\bf{\lambda }}^{\left( n \right)}})$;
			\STATE  Solve the  optimization problem about ${p_i}$ in the coalition;
			\STATE $p_i^{(n + 1)} = {\left( {p_i^{(n)} - {\delta ^{(n)}}{{\left. {\frac{{\partial {L^{(n)}}({\bf{P}},{{\bf{\lambda }}^{\left( n \right)}})}}{{\partial {p_i}}}} \right|}_{{p_i} = p_i^{(n)}}}} \right)^ + }$;
			\IF {$p_i^{(n + 1)} > P _{max}$}
			\STATE $p_i^{(n + 1)} = P_{max}$;
			\ENDIF
			\IF {${\delta ^{(n)}} > 1$}
			\STATE${\delta ^{(n + 1)}} = \frac{{{\delta ^{(n)}}}}{2}$;
			\ENDIF
			\STATE $\lambda _i^{(n + 1)} = {\left( {\lambda _i^{(n)} - {\mu ^{(n)}}{{\left( {{{\left. {\frac{{\partial {L^{(n)}}({{\bf{P}}^{(n + 1)}},{\bf{\lambda }})}}{{\partial {\lambda _i}}}} \right|}_{{\lambda _i} = \lambda _i^{(n)}}}} \right)}^ + }} \right)^ + },i = 1,...,K,K + 1$;
			\IF {${\mu ^{(n)}} > 1$}
			\STATE${\mu ^{(n + 1)}} = \frac{{{\mu ^{(n)}}}}{2}$;
			\ENDIF
			\IF {$\left| {R({{\bf{P}}^{(n + 1)}}) - R({{\bf{P}}^{(n)}})} \right| < \varepsilon $}
			\STATE ${{\bf{P}}^{\bf{*}}} = {{\bf{P}}^{(n + 1)}}$;
			\ELSE
			\STATE $n = n + 1$, and go to step 1;
			\ENDIF
		\end{algorithmic}
	\end{algorithm}
	
	The first is to initialize the Lagrange multiplier $\lambda _i^{(n)},i = 1,2,...,K,K + 1$, give the transmission power, and set the initial values of the gradient descent step size ${\delta ^{(0)}}$ and ${\mu ^{(0)}}$ of the power, and ${\lambda _i}$ are also given, and then use the DC programming to construct the above optimization problem, and then use the Taylor expansion method to approximate the original non-convex problem. In the second step, the Lagrangian unconstrained function corresponding to the convex problem is defined \cite{zhao2021joint}. For the optimization variable, the opposite direction of the first-order partial derivative is selected as the optimal direction of the gradient, and the selected step size decreases the step size as the number of iterations increases. For each iteration, the power of users is limited by the condition $0 \le {p_i} \le {P_{max}},\forall i = 1,2,...,K,K + 1$, and if $p_i^{(n + 1)} > P _{max}$ occurs, let $P = P _{max}$. The ${{\bf{P}}^{(n)}}$ obtained above is used as the initial value for the next iteration. From the eleventh step to solving the optimal ${\lambda _i}$, optimizing ${\lambda _i}$ needs to replace the optimal value of ${p_i}$ in the fourth to seventh steps. When the difference between the system sum rates before and after two iterations is less than the set threshold, the algorithm ends. The identified optimal solution power ${{\bf{P}}^{\bf{*}}}$ is obtained, otherwise, the next iteration is performed to solve ${{\bf{P}}^{(n + 1)}}$.
	
	The above algorithm is based on Taylor expansion, and through gradient descent, it continuously approaches the optimal solution of power and converges within a limited range to a level close to the optimal solution.
	
	\subsection{Local Discrete Phase Search Algorithm}\label{Sec5-3}
	Under the condition of fixed transmission power ${\bf{P}}$, the power constraint is removed, and the original problem about the phase $\Theta $ is still non-convex. $\Theta $ contains a series of discrete variables, and the usable range of each phase shift depends on the quantization bit $e$ of RIS. Considering the above reasons, we use the local search method in Algorithm \ref{alg3} to solve this problem.
	
	\begin{algorithm}[t]
		\caption{Local Discrete Phase Search Algorithm}
		\renewcommand{\algorithmicrequire}{\textbf{Initialization:}}
		\renewcommand{\algorithmicensure}{\textbf{Output:}}
		\label{alg3}
		\begin{algorithmic}[1]
			\REQUIRE
			Determine the quantization bits of RIS, and determine the region to which the user belongs;
			\ENSURE
			${\Theta ^*}$
			\FOR{$u=1$ to $M$}
			\FOR{${l_z}=1$ to $N$}
			\FOR{${l_y}=1$ to $N$}
			\STATE Allocate all possible values to ${\theta _{{l_z},{l_y}}}$ in turn, and under the premise of satisfying the SINR constraint, select the maximum sum rate for the current RIS-assisted user, and determine $\theta _{{l_z},{l_y}}^*$;
			\ENDFOR
			\ENDFOR
			\ENDFOR
		\end{algorithmic}
	\end{algorithm}
	
	When adjusting the phase of one of the RISs, the phases of the other $M - 1$ RISs remain unchanged, and the remaining ${N^2} - 1$ elements in the adjusted RIS board are also fixed. The ${\theta _{{l_z},{l_y}}}$ of each element traverses all its possible values and does not violate the SINR. Select the optimal value under the constraints until every element of RIS is traversed. In the end, there is an optimal solution $\theta _{{l_z},{l_y}}^*$ for each element, and the set of all $\theta _{{l_z},{l_y}}^*$ determines ${\Theta ^*}$, until the RIS of all assisted users satisfies the phase optimality, and Algorithm \ref{alg3} is obtained.
	
	\subsection{Sum Rate Maximization Algorithm}\label{Sec5-4}
	A system sum rate maximization algorithm is proposed based on this, which is described in Algorithm \ref{alg4}. First, initialize the system, set the transmission power of $D+C$ connections to the maximum transmission power of the corresponding frequency band, and randomly generate the phase shift matrix of each RIS. Then run the coalition game formation algorithm to determine the working mode of the D2D user, and then run the single coalition power allocation algorithm and the local discrete phase search algorithm in turn to solve the optimal power value in the current state and the phase optimal value of the RIS, and update them alternately until the algorithm converges. Between two consecutive iterations, the sum rate of the system is less than a certain threshold $\left| {{R^{(\rho  + 1)}} - {R^{(\rho )}}} \right| < \varepsilon $, which is regarded as algorithm convergence. Since the system bandwidth is relatively large, $\varepsilon {\rm{ = 1}}{{\rm{0}}^{\rm{3}}}$ is taken to represent the convergence threshold.
	
	\begin{algorithm}[t]
		\caption{Sum Rate Maximization Algorithm}
		\renewcommand{\algorithmicrequire}{\textbf{Initialization:}}
		\renewcommand{\algorithmicensure}{\textbf{Output:}}
		\label{alg4}
		\begin{algorithmic}[1]
			\REQUIRE
			$\varepsilon {\rm{ = 1}}{{\rm{0}}^{\rm{3}}}$, $\rho  = 0$, set quantization bits $e$, set RIS scale $N$, randomly generated $\Theta$, and randomly assign the working mode ${X_{c,d_{ini}}}$ of the D2D user, which is determined by the corresponding working mode $p_i^{(0)} = {P_{max}}$;
			\STATE Through Algorithm \ref{alg1}, the working mode of D2D users is determined based on the sum rate;
			\STATE Determine the transmission power ${{\bf{P}}^{\bf{*}}}$ of all users by Algorithm \ref{alg2};
			\STATE Update the optimal phase of RIS by Algorithm \ref{alg3};
			\IF {$\left| {{R^{(\rho  + 1)}} - {R^{(\rho )}}} \right| < \varepsilon $}
			\STATE ${R^*} = {R^{(\rho  + 1)}}$;
			\STATE {\textbf{Output:}} ${{\bf{P}}^{\bf{*}}}$, ${\Theta ^*}$ and ${X_{c,d}}$;
			\ELSE
			\STATE $\rho  = \rho  + 1$, back to line 1;
			\ENDIF
		\end{algorithmic}
	\end{algorithm}
	
	\subsection{Convergence, Feasibility and Complexity Analysis}\label{Sec5-5}
	\subsubsection{Convergence}\label{Sec5-5-1}
	The iterative computation of three subproblems is involved in the sum rate maximization algorithm. When $\Theta $ and ${\bf{P}}$ are fixed, ${X_{c,d}}$ is solved using Algorithm \ref{alg1}. For Algorithm \ref{alg1}, Starting from any initial coalitional structure ${F_{ini}}$, the proposed algorithm will converge to a final disjoint coalition after a series of switch operations. Through the preference defined in (\ref{eq20}), we find that each switch operation will either yield a new partition or switch existing partitions. So, the system will form at most $C + 1$ partitions as there is only $C$ cellular users plus one mm-wave band. Some coalitions may degenerate into the sets of very few D2D pairs, and even be emptied. Since the given number of partitions of D2D to set ${\mathcal{D}}$ is the Bell number \cite{saad2009coalitional}, we come to the conclusion that the sequence of swap operations always terminates and converges to the final partition, thus completing the proof of the convergence of the coalition game  formation algorithm. When $\Theta $ and ${X_{c,d}}$ are fixed, ${\bf{P}}$ is solved using Algorithm \ref{alg2}. For Algorithm \ref{alg2}, after \cite{chen2020reconfigurable} verification, as the number of iterations increases, the original objective function ${f_i}({\bf{P}})$ also decreases monotonically, and the convergence of the original problem is proved. Since the phase updated by Algorithm \ref{alg3} satisfies the condition $R({{\bf{P}}^*},{\Theta ^*}^{(\rho + 1)}) \ge R({{\bf{P}}^*},{\Theta^*}^{(\rho)})$, the objective function is non-decreasing.  In addition, the discrete phase is limited, and the transmission power is also limited by the upper and lower bounds, which make the problem of sum rate guaranteed. Therefore, we have explained the convergence of the sum rate maximization algorithm.
	
	\subsubsection{Feasibility}\label{Sec5-5-2}
	For problem (\ref{eq15}), the stability of the coalition game formation algorithm was studied in \cite{chen2018resource} by using hedonic games \cite{feng2016reliable}. It can prove that the final partition of the coalition game formation algorithm is Nash-stable. The proposed sub-problem in (\ref{eq16}) was proven not always feasible in \cite{chen2020reconfigurable} and is only  feasible under certain conditions. The sufficient conditions for a feasible solution of ${\bf{P}}$ are proposed and proven.
	
	\subsubsection{Complexity}\label{Sec5-5-3}
	The complexity of the sum rate maximization algorithm is not only related to the number of iterations to achieve convergence threshold $\varepsilon$, but also related to the complexity of the coalition game formation algorithm, single coalition power allocation algorithm, and local discrete phase search algorithm. For the working mode switch, the total iterations $N_{outter}$ was given. The  computational complexity of  Algorithm \ref{alg1} can be approximated as $O({N_{cg}})$.
	For the power allocation sub-problem, the complexity was created by the number of gradient updates and the optimization of transmission power of all coalitions in each gradient iteration. The number of cellular links and D2D links are $D+C$, and we denote the number of iterations of the gradient descent method as $N_{grad}$.
	For the phase shift sub-problem, when we change the phase of $\{{l_y},{l_z}\}$, the phases of the other ${N^2} - 1$ elements are also fixed. There are $M=8$ RISs involved in the system, each RIS has $N^2$ elements, and each element of each RIS has $2^e$ phase options. So, the complexity of this algorithm is $O(M*N^2*2^e)$. We get the complexity of the proposed sum rate maximization algorithm as $O(N_{outter}*({N_{cg}}+{N_{grad}}*(D+C)+M*N^2*2^e))$.
	
	\section{Performance Evaluation}\label{Sec6}
	
	\subsection{Simulation Setup}\label{Sec6-1}
	Regarding the simulation, since the channel model is highly correlated with the positions of the RIS, transmitter, and receiver, a three-dimensional Cartesian coordinate system is first established to represent the spatial position of each communication node in the system.
	
	\begin{figure}[!t]
		\begin{center}
			\includegraphics*[width=1\columnwidth,height=1.5in]{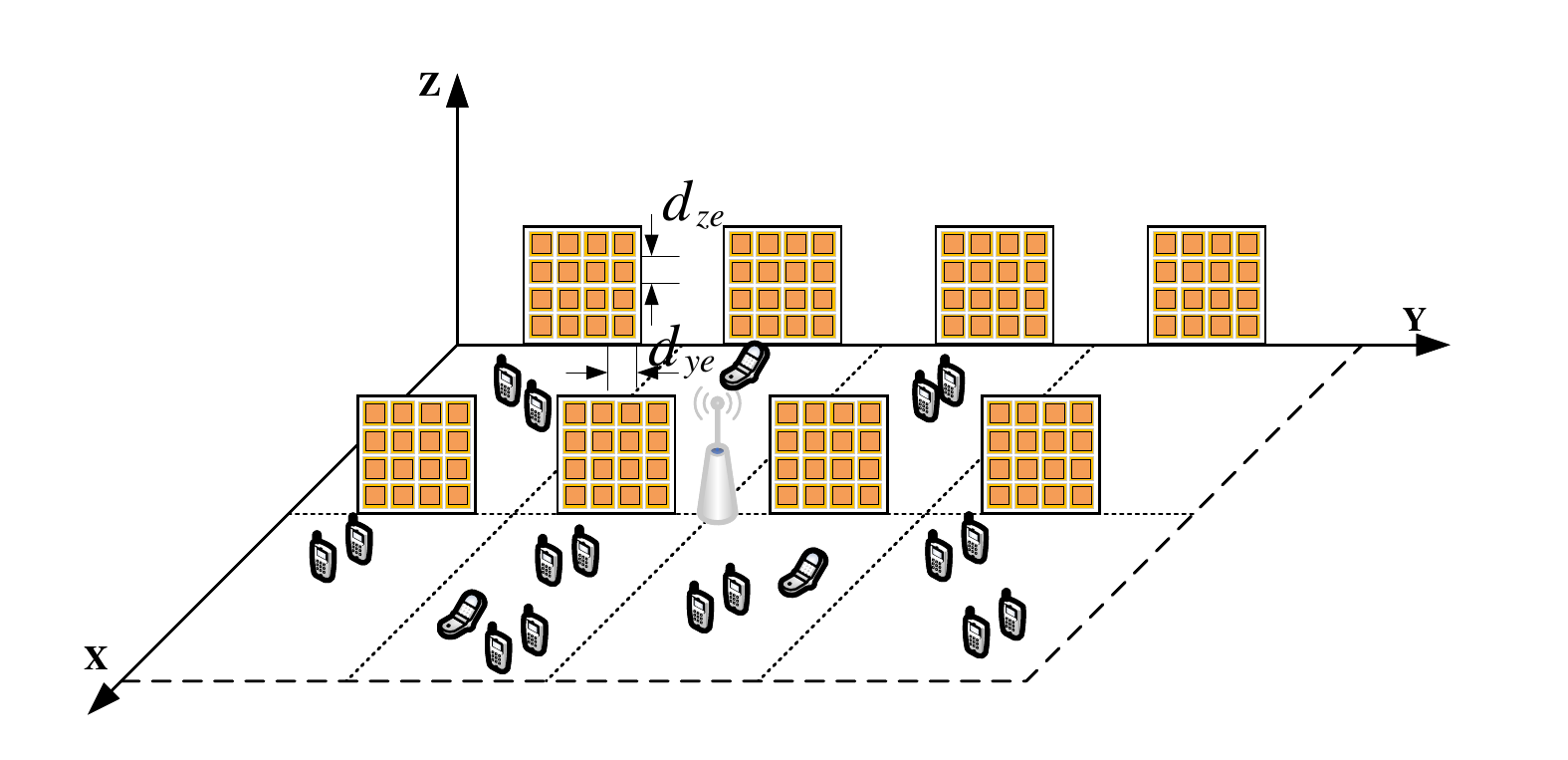}
		\end{center}
		\caption{3D coordinate map of RIS-assisted HCNs}
		\label{fig3.Network}
	\end{figure}
	
	As shown in Fig. \ref{Network}, multiple RIS blocks are placed in 3D space, one row is in the $Y{\rm{ - }}Z$ plane, and the other row is parallel to it. The coordinate of the lower left corner of the RIS closest to the origin is $(0,25,0)$, and the other RISs are arranged in $\Delta X = 50$ and $\Delta Y = 50$, with 4 RISs arranged in each row, with $m \in [1,4]$. Each RIS element has the upper right corner as the recorded coordinate $L{_{\{ l{_z},{l_y}\} }} = (0,25*m + l{_y}{d_{ye}},l{_z}{d_{ze}})$ or $L{_{\{ l{_z},{l_y}\} }} = (50,25*m + l{_y}{d_{ye}},l{_z}{d_{ze}})$, where the spacing between the Y and Z axes of adjacent elements of the same RIS is denoted as ${d_{ye}} = 0.005$ and ${d_{ze}} = 0.005$.
	
	In addition to this, assume that the entire system is deployed in a rectangular area on the $X - Y$ plane with vertices $(0,0,0)$, $(100,0,0)$, $(0,200,0)$, and $(100,200,0)$. The location of the BS is set to $(50,100,0)$, and the simulation is repeated 20 times for a fixed number of cellular users, and then the location results are averaged to obtain a more accurate location layout. According to the actual situation, set the maximum direct connection distance between D2D users to $10\sqrt 2 $ meters. Take link $i$ as an example, if ${L_{{t_i}}} = ({t_{ix}},{t_{iy}},0)$ and ${L_{{r_i}}} = ({r_{ix}},{r_{iy}},0)$. Based on the position of the RIS, the distance from ${t_i}$ to ${r_i}$ through the reflection of any element $\{ {l_y},{l_z}\}$ of the RIS is expressed as $D_{{t_i}}^{{l_y},{l_z}}$ and $D_{{l_y},{l_z}}^{{r_i}}$, and these two distances can be expressed as
	\begin{equation}\label{eq31}
		D_{{t_i}}^{{l_y},{l_z}} = \sqrt {{{({t_{ix}} - {l_x})}^2} + {{({t_{iy}} - {l_y}{d_{ye}})}^2} + {{({t_{iz}} - {l_z}{d_{ze}})}^2}},
	\end{equation}
	\begin{equation}\label{eq32}
		D_{{l_y},{l_z}}^{{r_i}} = \sqrt {{{({r_{ix}} - {l_x})}^2} + {{({r_{iy}} - {l_y}{d_{ye}})}^2} + {{({r_{iz}} - {l_z}{d_{ze}})}^2}}.
	\end{equation}
	
	Based on the above coordinate system, the four channel coefficients ${g_{{r_i},{t_i}}}$, $g_{{l_z},{l_y}}^{{r_i},{t_i}}$, ${h_{{r_i},{t_i}}}$ and $h_{{l_z},{l_y}}^{{r_i},{t_i}}$ involved in the channel model can be described in detail. For the reflection channel, the millimeter-wave reflection channel can be described as a Rician channel, including the LoS propagation component and the non-line-of-sight (NLoS) propagation component; the cellular-band reflection channel is described as a Rayleigh channel model with small-scale fading with propagation loss factor $n$, where the instantaneous channel tap is a function of time and space location, and ${\left| {{h_0}} \right|^2}$ is the second-order statistic of the channel, which is constant over the coverage of the BS. For link $i$, the two reflection channels are expressed as
	\begin{equation}\label{eq33}
		h_{{l_z},{l_y}}^{{r_i},{t_i}} = \sqrt {\frac{\beta }{{1 + \beta }}} \tilde h_{{l_z},{l_y}}^{{r_i},{t_i}} + \sqrt {\frac{1}{{1 + \beta }}} \hat h_{{l_z},{l_y}}^{{r_i},{t_i}},
	\end{equation}
	\begin{equation}\label{eq34}
		g_{{l_z},{l_y}}^{{r_i},{t_i}}{\rm{ = }}{h_0}\sqrt {{\alpha _i}{G_t}{G_r}{{(D_{{r_i}}^{{l_z},{l_y}} + D_{{l_z},{l_y}}^{{t_i}})}^{ - n}}},
	\end{equation}
	where $\beta  = 4$ is the Rice factor, $\tilde h_{{l_z},{l_y}}^{{r_i},{t_i}}$ is the LoS propagation component, and $\hat h_{{l_z},{l_y}}^{{r_i},{t_i}}$ is the NLoS propagation component, specifically expressed as (\ref{eq35}) and (\ref{eq36}).
	\begin{equation}\label{eq35}
		\tilde h_{{l_z},{l_y}}^{{r_i},{t_i}} = \sqrt {{\beta _0}{{(D_{{t_i}}^{{l_y},{l_z}} \cdot D_{{l_y},{l_z}}^{{r_i}})}^{ - \alpha }}} {e^{ - j\theta '}},
	\end{equation}
	\begin{equation}\label{eq36}
		\hat h_{{l_z},{l_y}}^{{r_i},{t_i}} = \sqrt {{\beta _0}{{(D_{{t_i}}^{{l_y},{l_z}} \cdot D_{{l_y},{l_z}}^{{r_i}})}^{ - \alpha '}}} \hat h_{NLoS,{l_z},{l_y}}^{{r_i},{t_i}},
	\end{equation}
	where ${\beta _0} =  - 61.3849 $ ${\rm{dB}}$ is the channel gain at a reference distance of 1 meter, $\theta '$ is a random phase in $[0,2\pi ]$, $\alpha  = 2.5$ is the loss index under LoS propagation, $\alpha ' = 3.6$ is the loss index under NLoS propagation, and $\hat h_{NLoS,{l_z},{l_y}}^{{r_i},{t_i}}\sim{\cal C}{\cal N}\left( {0,1} \right)$ is small-scale fading, and ${D_{{t_i}}^{{l_y},{l_z}} \cdot D_{{l_y},{l_z}}^{{r_i}}}$ is because of the far field model\cite{9206044}. Under the direct channel, the mm-wave band direct channel coefficient ${h_{{r_i},{t_i}}}$ can be expressed as
	\begin{equation}\label{eq37}
		{h_{{r_i},{t_i}}} = {h_i}\sqrt {{\beta _0}{{(D_{{t_i}}^{{r_i}})}^{ - \alpha }}} ,
	\end{equation}
	where ${h_i}$ is the small-scale fading that satisfies the ${\rm{Nakagami}} - {m_i}$ distribution, where the variable parameter are $\{ {m_i},{\omega _i}\} $, the fading depth parameter ${m_i} = 3$ , the average power of the fading signal ${\omega _i} = {1 \mathord{\left/{\vphantom {1 3}} \right.\kern-\nulldelimiterspace} 3}$ , and ${(D_{{t_i}}^{{r_i}})^{ - \alpha }}$ is the large-scale path loss, which depends on the distance $D_{{t_i}}^{{r_i}}$ between ${t_i}$ and ${r_i}$ to calculate.
	
	Under the cellular band direct channel, the coefficient ${g_{{r_i},{t_i}}}$ can be expressed as
	\begin{equation}\label{eq38}
		{g_{{r_i},{t_i}}}{\rm{ = }}{h_0}\sqrt {{G_t}{G_r}l_{{t_i},{r_i}}^{ - n}}  ,
	\end{equation}
	where ${h_0}$ is a complex Gaussian random variable with unity variance and zero mean, ${G_t}$ is the transmit antenna gain, ${G_r}$ is the receive antenna gain, ${l_{{t_i},{r_i}}}$ is the direct distance from ${t_i}$ to ${r_i}$, and $n=2$ is the path loss index.
	
	In this paper, the mm-wave band used in the simulation is the center frequency of 28 GHz, the bandwidth of 2160 MHz, the noise power spectral density of $-134 {\rm{dBm/MHz}}$, the maximum transmission power of 23 dBm, and the minimum SINR of all individuals is ${\gamma _{min}} = 5 {\rm{dB}}$. The bandwidth of the cellular band is 22 MHz, the noise power spectral density is $-174 {\rm{dBm/MHz}}$, the maximum transmission power is 20 dBm, and the obstacle density is ${\beta _1} = 0.01$. The simulation parameters are summarized in Table \ref{table1}.
	\begin{table}[t]
		\begin{center}
			\caption{SIMULATION PARAMETERS}
			\begin{tabular}{ccc}
				\hline
				Parameter & Symbol  & Value \\
				\hline
				Mm-wave bandwidth &  ${W_m}$  & 2160 MHz \\
				Cellular bandwidth & ${W_c}$ & 22 MHz\\
				Mm-wave noise spectral density & ${N_{0m}}$ & $-134 {\rm{dBm/MHz}}$ \\
				Cellular noise spectral density & ${N_{0c}}$ & $-174 {\rm{dBm/MHz}}$ \\
				Maximum power for mm-wave users  &  ${P_m}$   &  23 dBm \\
				Maximum power for cellular users &  ${P_c}$  & 20 dBm \\
				Cellular Path Loss Index &  $n$  & 2 \\
				Mm-wave LoS loss index &  $\alpha $  & 2.5 \\
				Mm-wave NLoS loss index &  $\alpha '$  & 3.6 \\
				Mm-wave channel gain &  ${\beta _0}$  & $ - 61.3849{\rm{dB}}$ \\
				Device antenna gain &   ${G_0}$ &    0.5 dBi \\
				BS antenna gain &   ${G_b}$ &    14 dBi \\
				Minimum signal-to-noise ratio &  ${\gamma _{min}}$ &  5 dB\\
				D2D user maximum distance  &   $r$   & $10\sqrt 2 m$ \\
				\hline
			\end{tabular}
			\label{table1}
		\end{center}
	\end{table}
	
	To demonstrate the superiority of the \textbf{Proposed algorithm} system performance, it is compared with the following algorithms:
	\begin{itemize}
		\item \textbf{Maximum Power Transmission (MP)}: In this scheme, both cellular users and D2D users adopt maximum power transmission and use the above-mentioned local discrete phase search algorithm to adjust the phase, and D2D users also use the  coalition  game formation algorithm to switch the working mode, which reduces power allocation compared with the proposed algorithm.
		\item \textbf{Random Phase Shift (RP)}: This scheme uses a set of feasible random phase RIS for assistance, keeps its phase unchanged, uses the maximum transmission power as the initial power, uses Algorithm \ref{alg1} to select the working mode of D2D users, and uses Algorithm \ref{alg2} to allocate power.
		\item \textbf{NonRIS}: This scheme does not use RIS for phase assistance, which means that the signal transmitted between the transmitter and the receiver only considers the direct channel. At the same time, the scheme also uses Algorithm \ref{alg2} and Algorithm \ref{alg3} for assistance.
		\item \textbf{Random Communication (NonCG)}: This scheme does not use the coalition game formation algorithm to switch the working mode of D2D users, adopts the method of randomly assigning the working mode of D2D users, and uses Algorithm \ref{alg2} for power allocation and Algorithm \ref{alg3} for phase assistance.
		\item \textbf{Full Mm-wave Communication(Fmm)}: This scheme does not use the coalition game formation algorithm to switch the working mode of D2D users. All D2D users work in the mm-wave frequency band, Algorithm \ref{alg2} is used for power allocation, and Algorithm \ref{alg3} is used for phase assistance.
	\end{itemize}
	
	\subsection{Performance Evalulation}\label{Sec6-2}
	Set $N{\rm{ = 4}}$, $e = 3$, and other parameters as in Table \ref{table1}, fix D2D users to 10 pairs and change the number of cellular users from 1 to 10. Six schemes sum rate performance curves are shown in Fig. \ref{ChangeCU}.
	\begin{figure}[!t]
		\begin{center}
			\includegraphics*[width=1\columnwidth,height=2.5in]{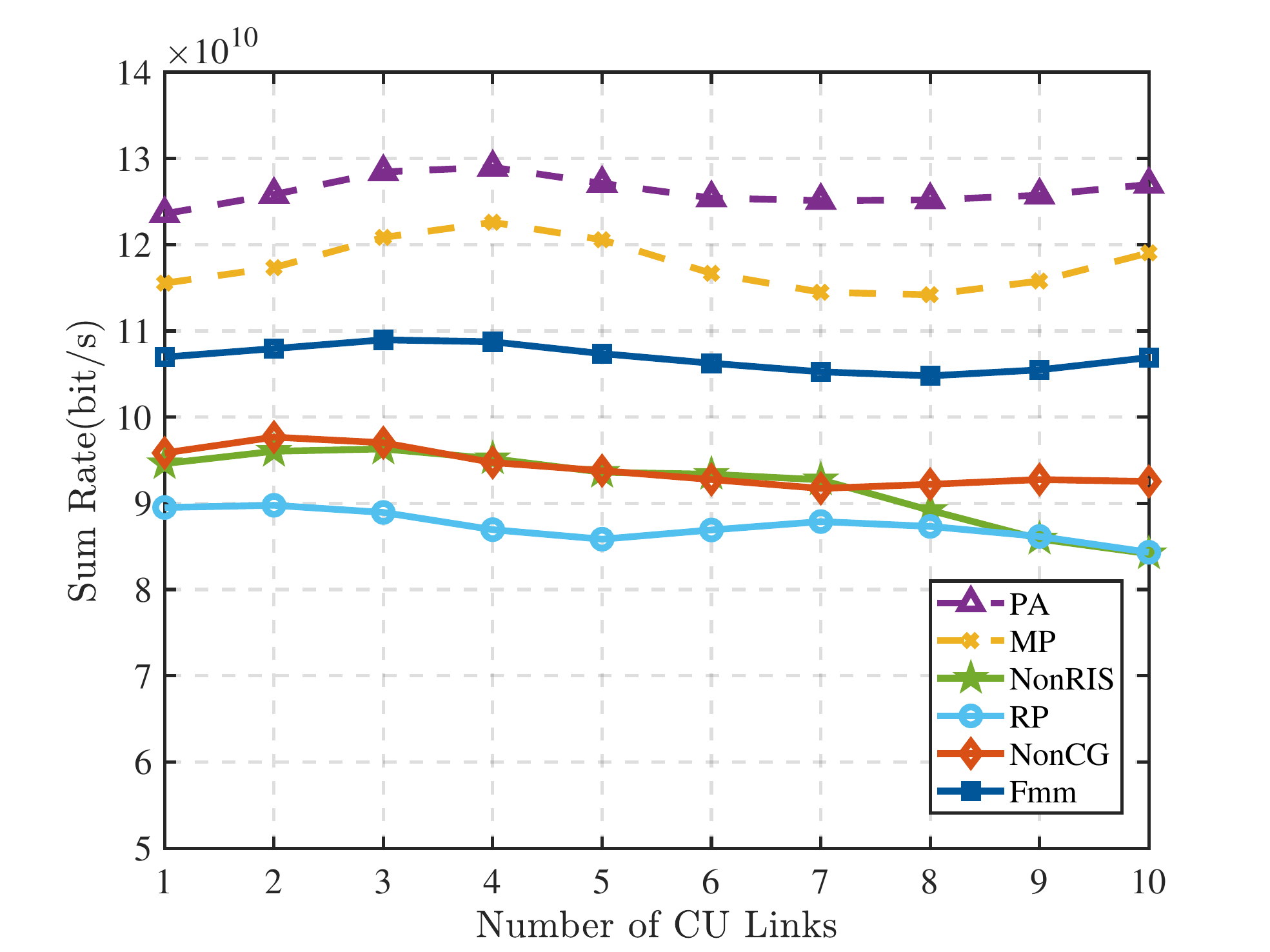}
		\end{center}
		\caption{System sum rate comparison chart of six different schemes when cellular users are different.} \label{fig4.ChangeCU}
	\end{figure}
	
	It can be observed from  Fig. \ref{ChangeCU} that with the increase of the number of cellular users, the total sum rate fluctuation range of the system is not large. This is because the mm-wave communication rate is much greater than the cellular communication rate, which leads the system to increase the number of D2D users in the mm-wave frequency band to maximize system sum rate. Comparing the six schemes, the proposed algorithm achieves the highest system sum rate. When the number of cellular users is 10, sum rates of the proposed algorithm (PA) are 3.9\% higher than the MP scheme, 35.7\% higher than the NonRIS scheme, 37.0\% higher than the RP scheme, 28.8\% higher than the NonCG scheme, and 15.6\% higher than the Fmm scheme.
	
	Set $N{\rm{ = 4}}$, $e = 3$, and other parameters as in Table \ref{table1}, fix CUs  to 5, and change the number of D2D pairs from 6 to 15. Six schemes' sum rate performance curves are shown in Fig. \ref{ChangeD2D}. It can be seen from Fig. \ref{ChangeD2D} that the sum rate of the system increases with the increase of D2D users, and the performance of the proposed algorithm is better than other comparable schemes.
	\begin{figure}[!t]
		\begin{center}
			\includegraphics*[width=1\columnwidth,height=2.5in]{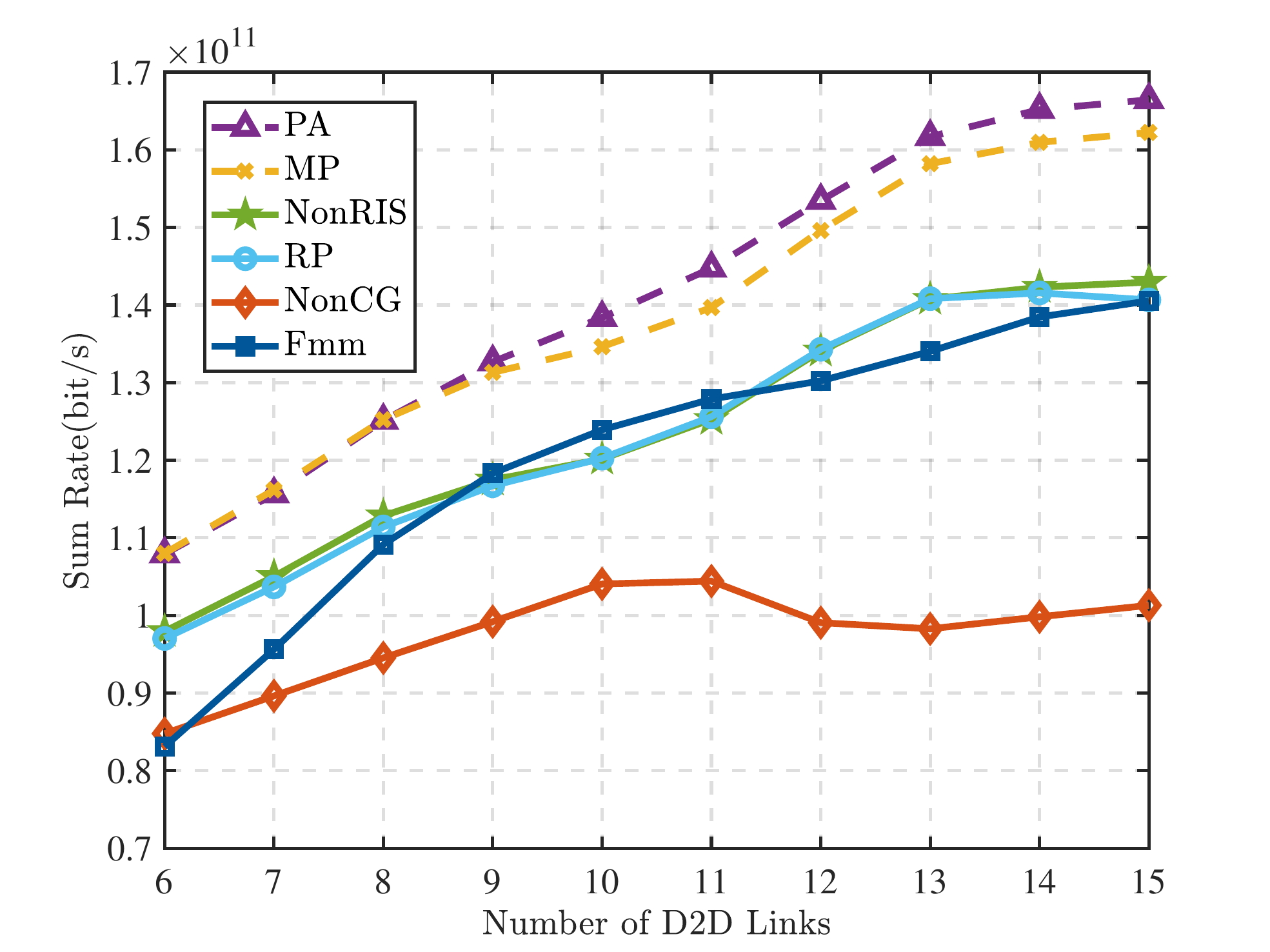}
		\end{center}
		\caption{System sum rate comparison diagram of six different schemes when changing D2D users} \label{fig5.ChangeD2D}
	\end{figure}
	
	When the number of D2D pairs is 15, the sum rate of PA is 1.8\% higher than the MP scheme, 13.7\% higher than the NonRIS scheme, 16.4\% higher than the RP scheme, 38.4\% higher than the NonCG scheme, and 16.4\% higher than the Fmm scheme. As the number of D2D users increases, the rate of random communication increases slowly. This is because there are more frequency bands for cellular users, which makes it easier to allocate cellular frequency bands. With the increase of D2D users, the system can improve the spectrum utilization rate, but the interference management capability is limited, so the system performance is also limited and the rate growth rate slows down.
	
	Set $e = 3$, and other parameters as in Table \ref{table1}, fix CUs to 5 , fix D2D pairs to 10, and change the number of RIS elements from ${{\rm{2}}^2}$ to ${{\rm{2}}^{14}}$. As illustrated in Fig. \ref{ChangeN}, the sum rate of the proposed algorithm system increases with the increase of $N$. When $N=14$, the sum rate of PA is 1.7\% higher than the MP scheme, 22.6\% higher than the NonRIS scheme, 25.6\% higher than the RP scheme, 18.9\% higher than the  NonCG scheme, and 11.4\% higher than the Fmm scheme.
	\begin{figure}[!t]
		\begin{center}
			\includegraphics*[width=1\columnwidth,height=2.5in]{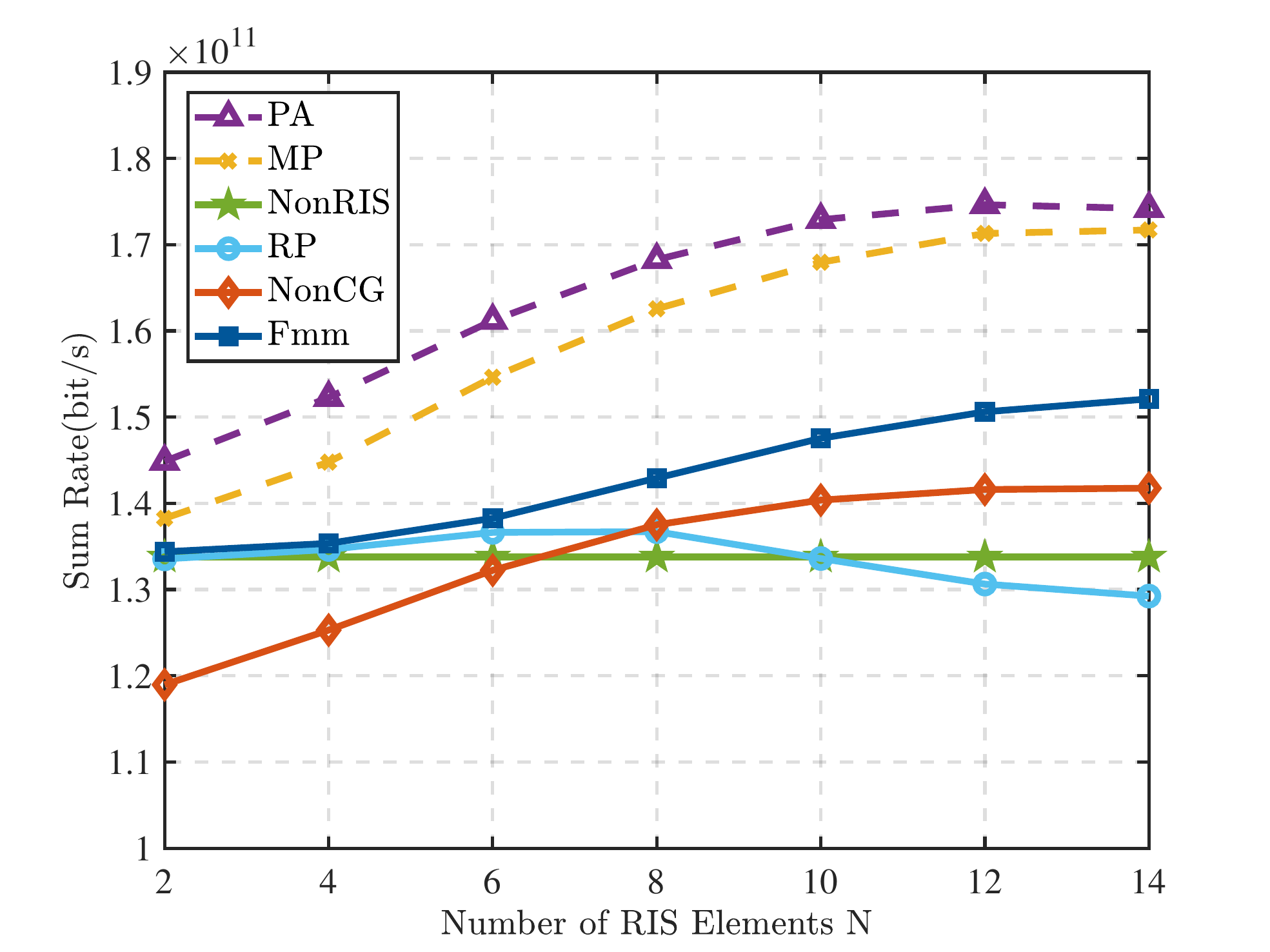}
		\end{center}
		\caption{Comparison chart of system sum rate of six different schemes when RIS element N is different} \label{fig6.ChangeN}
	\end{figure}
	
	Using RIS for beamforming reduces unwanted co-channel interference. It is worth noting that the sum rate growth of PA and MP schemes slows down slightly when the number of elements is large. This is because the increase of RIS elements not only increases the number of useful signals, but also increases the paths of interfering signals. Concomitantly, the increase of the sum rate brought by continuing to increase the number of RIS elements will also slow down for the NonCG and Fmm schemes. For the random phase scheme (RP), there is no constructive alignment of the reflected beams, the increase of RIS elements will lead to increased interference, and the sum rate of the system depends on the quality of each random, so the system sum rate cannot grow steadily. For the random communication scheme (NonCG), due to the lack of resource allocation methods, the utilization of mm-wave resources is incomplete, and the system sum rate is mainly determined by the utilization rate of the millimeter-band, so it is far from the proposed algorithm (PA). As the number of components $N$ increases, the sum rate exhibited by the RP scheme gradually decreases, again due to the lack of interference management by the RIS.
	\begin{figure}[!t]
		\begin{center}
			\includegraphics*[width=1\columnwidth,height=2.5in]{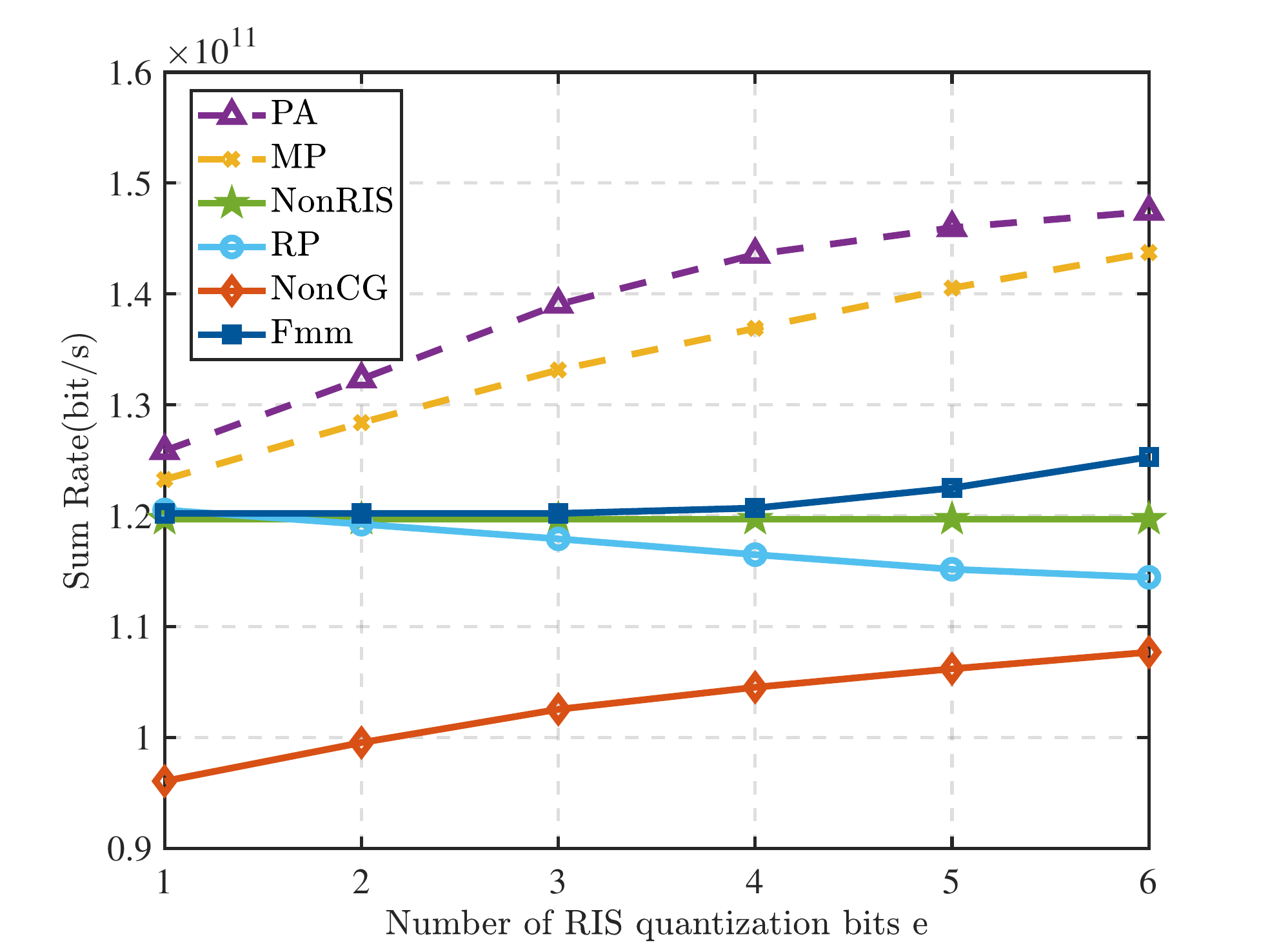}
		\end{center}
		\caption{System sum rate comparison diagram of six different schemes when the quantization bit e is different} \label{fig7.ChangeE}
	\end{figure}
	
	Set $N = 4$, and other parameters as in Table \ref{table1}, fix CUs to 5 , fix D2D pairs to 10, and change the number of RIS quantization bits $e$ from 1 to 6.  As illustrated in Fig. \ref{ChangeE}, it can be obtained that under such parameter settings, the number of quantization bits has a more prominent influence on the signal superposition in the reflection link.
	
	There are several more important observations from Fig. \ref{ChangeE}. First, the proposed algorithm (PA) significantly outperforms the other 5 benchmark algorithms. Secondly, the systems sum rates of PA, MP, NonCG, and Fmm schemes increase with the increase of quantization bits $e$. As the number of quantization bits increases, the growth sum rate of several of these schemes slows down when the quantization bits reach 5 and 6. This indicates that the system performance tends to be saturated when the number of quantization bits exceeds 4. In addition to this, the system sum rate of the RP scheme shows signs of decay with increasing quantization bits. The reason for this phenomenon is that the distances of the reflected links that generate the useful signal and the interference signal are random, and the phase is also randomly generated, so the useful reflected signal is not necessarily stronger than the interference signal. At the same time, the range of the constructive phase is relatively small, and increasing the quantization bit will lead to a smaller probability of random to an optimized phase, and interference will prevail, resulting in a decrease in the system and rate. From the perspective of quantitative analysis, when e=4, the sum rate of PA is 1.3\% higher than the MP scheme, 19.5\% higher than the NonRIS scheme, 23.3\% higher than the RP scheme, 26.6\% higher than the NonCG scheme, and 13.2\% higher than the Fmm scheme.
	
	\section{Conclusion}\label{Sec7}
	In this paper, we study the D2D communication of the HCN combining the uplink mm-wave with the traditional cellular network. We propose the corresponding system model, and express different channel models for the channels of different bands and different propagation modes. We expound on resource allocation problem between multiple pairs of D2D mm-wave and cellular bands from the perspective of game theory and propose a coalition game formation algorithm based on game theory to switch the working mode of D2D devices in D2D communication based on HCNs. At the same time, the RIS technology is introduced to assist the channel, enhance the useful signal, and reduce co-channel interference. A single coalition power allocation algorithm and local discrete phase search algorithm based on the coalition are designed to maximize the sum rate of the system under the condition of ensuring QoS. Finally, the mentioned algorithm is integrated and iterated, and the superiority of the designed algorithm is verified in the simulation results. It was also observed that RIS can convert a poorly conditioned channel into a well-conditioned channel by providing multiple controllable reflection signals based on the number of RIS elements. In summary, the proposed algorithm shows the superiority of the rate compared with other schemes.

	\bibliographystyle{IEEEtran}
	\bibliography{references}

\begin{thebibliography}{10}
\providecommand{\url}[1]{#1}
\csname url@samestyle\endcsname
\providecommand{\newblock}{\relax}
\providecommand{\bibinfo}[2]{#2}
\providecommand{\BIBentrySTDinterwordspacing}{\spaceskip=0pt\relax}
\providecommand{\BIBentryALTinterwordstretchfactor}{4}
\providecommand{\BIBentryALTinterwordspacing}{\spaceskip=\fontdimen2\font plus
\BIBentryALTinterwordstretchfactor\fontdimen3\font minus
  \fontdimen4\font\relax}
\providecommand{\BIBforeignlanguage}[2]{{%
\expandafter\ifx\csname l@#1\endcsname\relax
\typeout{** WARNING: IEEEtran.bst: No hyphenation pattern has been}%
\typeout{** loaded for the language `#1'. Using the pattern for}%
\typeout{** the default language instead.}%
\else
\language=\csname l@#1\endcsname
\fi
#2}}
\providecommand{\BIBdecl}{\relax}
\BIBdecl

\bibitem{forecast2019cisco}
``Cisco visual networking index: global mobile data traffic forecast update,
  2017--2022,'' pp. 1--33, Feb. 2019.

\bibitem{ecma60ghz}
H.~R. ECMA, ``Ghz phy, mac and pals, std. ecma-387, retrieved aug., 2, 2014,''
  Aug. 60.

\bibitem{5284444}
``Ieee standard for information technology-- local and metropolitan area
  networks-- specific requirements-- part 15.3: Amendment 2:
  Millimeter-wave-based alternative physical layer extension,'' \emph{IEEE Std
  802.15.3c-2009 (Amendment to IEEE Std 802.15.3-2003)}, pp. 1--200, Oct. 2009.

\bibitem{singh2011interference}
S.~Singh, R.~Mudumbai, and U.~Madhow, ``Interference analysis for highly
  directional 60-ghz mesh networks: The case for rethinking medium access
  control,'' \emph{IEEE/ACM Transactions on Networking}, vol.~19, no.~5, pp.
  1513--1527, Oct. 2011.

\bibitem{9779354}
J.~Li, Y.~Niu, H.~Wu, B.~Ai, S.~Chen, Z.~Feng, Z.~Zhong, and N.~Wang,
  ``Mobility support for millimeter wave communications: Opportunities and
  challenges,'' \emph{IEEE Communications Surveys \& Tutorials}, vol.~24,
  no.~3, pp. 1816--1842, Third quarter 2022.

\bibitem{huang2019reconfigurable}
C.~Huang, A.~Zappone, G.~C. Alexandropoulos, M.~Debbah, and C.~Yuen,
  ``Reconfigurable intelligent surfaces for energy efficiency in wireless
  communication,'' \emph{IEEE Transactions on Wireless Communications},
  vol.~18, no.~8, pp. 4157--4170, Aug. 2019.

\bibitem{wu2019beamforming}
Q.~Wu and R.~Zhang, ``Beamforming optimization for wireless network aided by
  intelligent reflecting surface with discrete phase shifts,'' \emph{IEEE
  Transactions on Communications}, vol.~68, no.~3, pp. 1838--1851, Mar. 2019.

\bibitem{zhang2021reconfigurable}
H.~Zhang, B.~Di, L.~Song, and Z.~Han, \emph{Reconfigurable Intelligent
  Surface-Empowered 6G}.\hskip 1em plus 0.5em minus 0.4em\relax Springer, 2021.

\bibitem{tang2020wireless}
W.~Tang, M.~Z. Chen, X.~Chen, J.~Y. Dai, Y.~Han, M.~Di~Renzo, Y.~Zeng, S.~Jin,
  Q.~Cheng, and T.~J. Cui, ``Wireless communications with reconfigurable
  intelligent surface: Path loss modeling and experimental measurement,''
  \emph{IEEE Transactions on Wireless Communications}, vol.~20, no.~1, pp.
  421--439, Jan. 2020.

\bibitem{5519540}
N.~Vucic, S.~Shi, and M.~Schubert, ``Dc programming approach for resource
  allocation in wireless networks,'' in \emph{8th International Symposium on
  Modeling and Optimization in Mobile, Ad Hoc, and Wireless Networks}, Avignon,
  France, May/Jun 2010, pp. 380--386.

\bibitem{ramezani2017joint}
A.~Ramezani-Kebrya, M.~Dong, B.~Liang, G.~Boudreau, and S.~H. Seyedmehdi,
  ``Joint power optimization for device-to-device communication in cellular
  networks with interference control,'' \emph{IEEE Transactions on Wireless
  Communications}, vol.~16, no.~8, pp. 5131--5146, Aug. 2017.

\bibitem{zhao2015social}
Y.~Zhao, Y.~Li, Y.~Cao, T.~Jiang, and N.~Ge, ``Social-aware resource allocation
  for device-to-device communications underlaying cellular networks,''
  \emph{IEEE Transactions on Wireless Communications}, vol.~14, no.~12, pp.
  6621--6634, Dec. 2015.

\bibitem{chen2018resource}
Y.~Chen, B.~Ai, Y.~Niu, K.~Guan, and Z.~Han, ``Resource allocation for
  device-to-device communications underlaying heterogeneous cellular networks
  using coalitional games,'' \emph{IEEE Transactions on Wireless
  Communications}, vol.~17, no.~6, pp. 4163--4176, Jun. 2018.

\bibitem{wang2013position}
H.~Wang, K.~Xia, and X.~Chu, ``On the position-based resource-sharing for
  device-to-device communications underlaying cellular networks,'' in
  \emph{IEEE/CIC International Conference on Communications in China (ICCC)},
  Xi'an, China, Aug. 2013, pp. 135--140.

\bibitem{feng2017effective}
Z.~Feng, Z.~Feng, and T.~A. Gulliver, ``Effective small social community aware
  d2d resource allocation underlaying cellular networks,'' \emph{IEEE Wireless
  Communications Letters}, vol.~6, no.~6, pp. 822--825, Dec. 2017.

\bibitem{8283645}
B.~Zhang, X.~Mao, J.-L. Yu, and Z.~Han, ``Resource allocation for 5g
  heterogeneous cloud radio access networks with d2d communication: A matching
  and coalition approach,'' \emph{IEEE Transactions on Vehicular Technology},
  vol.~67, no.~7, pp. 5883--5894, Jul. 2018.

\bibitem{7890452}
S.~M.~A. Kazmi, N.~H. Tran, W.~Saad, Z.~Han, T.~M. Ho, T.~Z. Oo, and C.~S.
  Hong, ``Mode selection and resource allocation in device-to-device
  communications: A matching game approach,'' \emph{IEEE Transactions on Mobile
  Computing}, vol.~16, no.~11, pp. 3126--3141, Nov. 2017.

\bibitem{9849051}
H.~Du, D.~Niyato, Y.-A. Xie, Y.~Cheng, J.~Kang, and D.~I. Kim, ``Performance
  analysis and optimization for jammer-aided multiantenna uav covert
  communication,'' \emph{IEEE Journal on Selected Areas in Communications},
  vol.~40, no.~10, pp. 2962--2979, Oct. 2022.

\bibitem{deng2018millimeter}
N.~Deng, M.~Haenggi, and Y.~Sun, ``Millimeter-wave device-to-device networks
  with heterogeneous antenna arrays,'' \emph{IEEE Transactions on
  Communications}, vol.~66, no.~9, pp. 4271--4285, Sept. 2018.

\bibitem{chen2019resource}
Y.~Chen, B.~Ai, Y.~Niu, R.~He, Z.~Zhong, and Z.~Han, ``Resource allocation for
  device-to-device communications in multi-cell multi-band heterogeneous
  cellular networks,'' \emph{IEEE Transactions on Vehicular Technology},
  vol.~68, no.~5, pp. 4760--4773, May 2019.

\bibitem{wu2019intelligent}
Q.~Wu and R.~Zhang, ``Intelligent reflecting surface enhanced wireless network
  via joint active and passive beamforming,'' \emph{IEEE Transactions on
  Wireless Communications}, vol.~18, no.~11, pp. 5394--5409, Nov. 2019.

\bibitem{zhang2020reconfigurable}
H.~Zhang, B.~Di, L.~Song, and Z.~Han, ``Reconfigurable intelligent surfaces
  assisted communications with limited phase shifts: How many phase shifts are
  enough?'' \emph{IEEE Transactions on Vehicular Technology}, vol.~69, no.~4,
  pp. 4498--4502, Apr. 2020.

\bibitem{wu2019towards}
Q.~Wu and R.~Zhang, ``Towards smart and reconfigurable environment: Intelligent
  reflecting surface aided wireless network,'' \emph{IEEE Communications
  Magazine}, vol.~58, no.~1, pp. 106--112, Jan. 2020.

\bibitem{hou2020reconfigurable}
T.~Hou, Y.~Liu, Z.~Song, X.~Sun, Y.~Chen, and L.~Hanzo, ``Reconfigurable
  intelligent surface aided noma networks,'' \emph{IEEE Journal on Selected
  Areas in Communications}, vol.~38, no.~11, pp. 2575--2588, Nov. 2020.

\bibitem{li2020reconfigurable}
S.~Li, B.~Duo, X.~Yuan, Y.-C. Liang, and M.~Di~Renzo, ``Reconfigurable
  intelligent surface assisted uav communication: Joint trajectory design and
  passive beamforming,'' \emph{IEEE Wireless Communications Letters}, vol.~9,
  no.~5, pp. 716--720, May 2020.

\bibitem{zappone2020optimal}
A.~Zappone, M.~Di~Renzo, X.~Xi, and M.~Debbah, ``On the optimal number of
  reflecting elements for reconfigurable intelligent surfaces,'' \emph{IEEE
  Wireless Communications Letters}, vol.~10, no.~3, pp. 464--468, Mar. 2021.

\bibitem{pan2020multicell}
C.~Pan, H.~Ren, K.~Wang, W.~Xu, M.~Elkashlan, A.~Nallanathan, and L.~Hanzo,
  ``Multicell mimo communications relying on intelligent reflecting surfaces,''
  \emph{IEEE Transactions on Wireless Communications}, vol.~19, no.~8, pp.
  5218--5233, Aug. 2020.

\bibitem{zheng2019intelligent}
B.~Zheng and R.~Zhang, ``Intelligent reflecting surface-enhanced ofdm: Channel
  estimation and reflection optimization,'' \emph{IEEE Wireless Communications
  Letters}, vol.~9, no.~4, pp. 518--522, Apr. 2020.

\bibitem{wei2021channel}
L.~Wei, C.~Huang, G.~C. Alexandropoulos, C.~Yuen, Z.~Zhang, and M.~Debbah,
  ``Channel estimation for ris-empowered multi-user miso wireless
  communications,'' \emph{IEEE Transactions on Communications}, vol.~69, no.~6,
  pp. 4144--4157, Jun. 2021.

\bibitem{wei2020parallel}
L.~Wei, C.~Huang, G.~C. Alexandropoulos, and C.~Yuen, ``Parallel factor
  decomposition channel estimation in ris-assisted multi-user miso
  communication,'' in \emph{IEEE 11th sensor array and multichannel signal
  processing workshop (SAM)}, Hangzhou, China, Jun. 2020.

\bibitem{jung2016connectivity}
H.~Jung and I.-H. Lee, ``Connectivity analysis of millimeter-wave
  device-to-device networks with blockage,'' \emph{International Journal of
  Antennas and Propagation}, vol. 2016, Nov. 2016.

\bibitem{von2007theory}
J.~Von~Neumann and O.~Morgenstern, ``Theory of games and economic behavior,''
  in \emph{Theory of games and economic behavior}.\hskip 1em plus 0.5em minus
  0.4em\relax Princeton university press, 2007.

\bibitem{zhao2021joint}
D.~Zhao, H.~Lu, Y.~Wang, H.~Sun, and Y.~Gui, ``Joint power allocation and user
  association optimization for irs-assisted mmwave systems,'' \emph{IEEE
  Transactions on Wireless Communications}, vol.~21, no.~1, pp. 577--590, Jan.
  2022.

\bibitem{saad2009coalitional}
W.~Saad, Z.~Han, M.~Debbah, A.~Hjorungnes, and T.~Basar, ``Coalitional game
  theory for communication networks,'' \emph{IEEE Signal Processing Magazine},
  vol.~26, no.~5, pp. 77--97, Sept. 2009.

\bibitem{chen2020reconfigurable}
Y.~Chen, B.~Ai, H.~Zhang, Y.~Niu, L.~Song, Z.~Han, and H.~Vincent~Poor,
  ``Reconfigurable intelligent surface assisted device-to-device
  communications,'' \emph{IEEE Transactions on Wireless Communications},
  vol.~20, no.~5, pp. 2792--2804, May 2021.

\bibitem{feng2016reliable}
R.~Feng, T.~Li, Y.~Wu, and N.~Yu, ``Reliable routing in wireless sensor
  networks based on coalitional game theory,'' \emph{Iet Communications},
  vol.~10, no.~9, pp. 1027--1034, Jun. 2016.

\bibitem{9206044}
W.~Tang, M.~Z. Chen, X.~Chen, J.~Y. Dai, Y.~Han, M.~Di~Renzo, Y.~Zeng, S.~Jin,
  Q.~Cheng, and T.~J. Cui, ``Wireless communications with reconfigurable
  intelligent surface: Path loss modeling and experimental measurement,''
  \emph{IEEE Transactions on Wireless Communications}, vol.~20, no.~1, pp.
  421--439, Jan. 2021.

\end{thebibliography}

\end{document}